\documentclass[twocolumn]{aastex701}
\usepackage{xcolor}
\usepackage{subcaption}

\usepackage{amsmath}
\usepackage{tablefootnote}
\usepackage{multirow}
\usepackage{placeins}
\usepackage{comment}
\usepackage{hyperref}
\usepackage{makecell}
\usepackage{soul}

\begin{document}

\shorttitle{Effects of Atmosphere Mismatch on NICER Radius Inference}
\shortauthors{Isiah M. Holt et al.}

\graphicspath{{./}{figures/}}

\title{Systematic Effects of Hydrogen and Helium Atmosphere Mismatch on Radius Inference in PSR~J0740+6620-like Synthetic NICER Data}
\author[0000-0002-3097-942X]{Isiah M. Holt}
\affiliation{Department of Astronomy, University of Maryland, College Park, MD 20742-2421, USA}
\affiliation{NASA Goddard Space Flight Center, Greenbelt, MD, USA}
\email{imholt@umd.edu}

\author[0000-0002-2666-728X]{M. Coleman Miller}
\affiliation{Department of Astronomy, University of Maryland, College Park, MD 20742-2421, USA}
\affiliation{Joint Space-Science Institute, University of Maryland, College Park, MD 20742-2421, USA}
\email{mcmiller@umd.edu}

\author[0000-0001-6157-6722]{Alexander J. Dittmann}
\affiliation{Institute for Advanced Study, 1 Einstein Drive, Princeton, NJ 08540, USA}
\altaffiliation{NASA Einstein Fellow}
\email{dittmann@ias.edu}

\author[0000-0002-3862-7402]{Frederick K. Lamb}
\affiliation{Illinois Center for Advanced Studies of the Universe and Department of Physics, University of Illinois at Urbana-Champaign, 1110 West Green Street, Urbana, IL 61801-3080, USA}
\affiliation{Department of Astronomy, University of Illinois at Urbana-Champaign, 1002 West Green Street, Urbana, IL 61801-3074, USA}
\email{fkl@illinois.edu}

\begin{abstract}
Constraints on neutron star radii provide insight into the properties of the cold, dense matter in their interiors. Previous studies using synthetic Neutron star Interior Composition Explorer (NICER) pulse waveform data have demonstrated that radius inferences derived therefrom are robust against several classes of modeling systematics.  Here we explore the consequences of assuming the wrong atmospheric composition, using synthetic data based on the $\sim 2.1~M_\odot$ pulsar PSR~J0740$+$6620.  We find that the assumption of a hydrogen atmosphere when the synthetic data assumed a helium atmosphere, or vice versa, produces little bias in the inferred radius at the spot-to-background ratio of the actual PSR~J0740$+$6620 data.  However, when we increase the spot-to-background ratio by a factor of $\sim20$ while keeping the total number of counts fixed at the observed $\sim 5.5\times 10^5$, we find that composition mismatch can produce significantly biased radius estimates while remaining hidden inside a fit with statistically acceptable residuals.  Even in these cases, the Bayesian evidence consistently identifies the correct atmospheric model. Our findings reinforce the importance of using the Bayesian evidence for model comparison and goodness-of-fit tests.
\end{abstract}


\section{Introduction}\label{sec:intro}

The cores of neutron stars consist of matter that is theorized to be catalyzed to its ground state at densities up to several times nuclear saturation density ($n_s \approx 0.16$~fm$^{-3}$, equivalent to a mass density $\rho_s \approx 2.6 \times 10^{14}$~g~cm$^{-3}$).  Laboratories are unable to replicate the high densities, low temperatures, and extreme neutron--proton asymmetries present in neutron star cores.  Thus, observations of neutron stars provide unique insights into the nature of dense matter.  Most of the macroscopic properties of a neutron star, such as its mass, radius, and tidal deformability, depend on the equation of state (EOS) of the matter in its interior, i.e., the pressure as a function of the energy density.  Over the past decade, numerous observations have helped constrain the EOS.  These include measurements of high neutron star masses such as PSR~J1614$-$2230 ($\rm{M} = 1.97 \pm 0.04~\rm{M}_\odot$, \citealt{2010Natur.467.1081D}), PSR~J0348$+$0432 ($\rm{M} = 1.806 \pm0.037~\rm{M}_\odot$, \citealt{2024arXiv241202850S}), PSR~J0740$+$6620 ($\rm{M} = 2.08 \pm 0.07~\rm{M}_\odot$, \citealt{2021ApJ...915L..12F}), and PSR~J0952$-$0607 ($\rm{M} = 2.35 \pm 0.17~\rm{M}_\odot$, \citealt{2022ApJ...934L..17R}). They also include constraints on the tidal deformability of neutron stars from gravitational-wave observations of the binary neutron star merger GW170817 (\citealt{2017PhRvL.119p1101A}; \citealt{2018PhRvL.121p1101A}; \citealt{2018PhRvL.121i1102D}), and estimates of neutron star radii from thermal X-ray observations.  Among these, measurements of the radii of neutron stars with precisely determined masses are especially valuable.

Fits of nonmagnetic helium model atmospheres to the spectra of cooling neutron stars in quiescent low-mass X-ray binary systems (qLMXBs;~\citealt{2011ApJ...732...88G,2013ApJ...772....7G,2014ApJ...784..123L,2016ApJ...831..184B}) commonly yield radius estimates $\sim 50\%$ larger than fits of nonmagnetic hydrogen model atmospheres to the same data (\citealt{2012MNRAS.423.1556S}; \citealt{2013ApJ...764..145C}; \citealt{2014MNRAS.444..443H}; see also \citealt{2016EPJA...52...63M} for a review). The origin of this discrepancy could lie in a spectral shape degeneracy.  Hydrogen and helium atmospheres produce emergent spectra of different hardness at the same effective temperature, so fitting a hydrogen atmosphere model to helium atmosphere data (or vice versa) forces a trade-off between the inferred temperature and emitting area, and hence the radius. The soft X-ray spectra of these sources are roughly thermal and lack strong composition-dependent features, so the data are often insufficient to distinguish between a smaller, hotter neutron star with one composition and a larger, cooler neutron star with another (see, e.g., \citealt{2016EPJA...52...63M}). Similarly, hydrogen and carbon model atmospheres can give equally good fits to the spectra of isolated cooling neutron stars while returning very different radii (see, e.g., \citealt{2015A&A...573A..53K}). In these spectral fitting methods, the composition ambiguity is difficult to resolve because the data lack the spectral features or independent geometric constraints needed to distinguish atmospheric models.

The Neutron Star Interior Composition Explorer (NICER; \citealt{2012SPIE.8443E..13G}) conducts phase- and energy-resolved measurements of the thermal X-ray pulses produced by some rotating neutron stars, which can be used to constrain the masses and radii of these stars. Mass and radius constraints have been derived from NICER data, sometimes supplemented with X-ray Multi-Mirror (XMM-Newton) data, for several neutron stars. These include the $\sim 1.4~M_\odot$ pulsar PSR~J0030$+$0451 (\citealt{2019ApJ...887L..24M}; \citealt{2019ApJ...887L..21R,2024ApJ...961...62V,2026ApJ..1005..201K}), the $\sim 2.1~M_\odot$ pulsar PSR~J0740$+$6620 (\citealt{2021ApJ...918L..28M}; \citealt{2021ApJ...918L..27R}; \citealt{2022ApJ...941..150S}; \citealt{2024ApJ...974..295D}; \citealt{2024ApJ...974..294S}; this pulsar is especially important because of its high mass and thus high central density), the $\sim 1.4~M_\odot$ pulsar PSR~J0437--4715 (\citealt{2024ApJ...971L..20C,2026ApJ..1000L..48M}), the $\sim 1.04~M_\odot$ pulsar PSR~1231--1411 (\citealt{2024ApJ...976...58S}; \citealt{2025ApJ...981...99Q}), the $\sim 1.4~M_\odot$ pulsar PSR~J0164--3329 (\citealt{2025ApJ...995...60M}), and the $\sim 1.8~M_\odot$ pulsar PSR~J2124--3358 (\citealt{2026arXiv260703721G}). In all such analyses, the atmosphere of the neutron star plays an important role, as it determines the angular distribution of the thermal X-ray radiation leaving the stellar surface.

In pulse-profile modeling (PPM), the predicted X-ray pulse waveform depends not only on the stellar compactness, observer geometry, and emission region properties, but also on the beaming of the emergent radiation, which is influenced by the atmospheric composition (see, e.g., \citealt{1983ApJ...274..846P}; \citealt{2003MNRAS.343.1301P}; \citealt{2013ApJ...776...19L}; \citealt{2015ApJ...808...31M}). Published PPM analyses of NICER data have predominantly assumed partially and/or fully ionized hydrogen atmospheres when modeling the X-ray emission (see, e.g., \citealt{2019ApJ...887L..24M}; \citealt{2019ApJ...887L..21R}; \citealt{2021ApJ...918L..28M}; \citealt{2021ApJ...918L..27R}; \citealt{2024ApJ...974..294S}; \citealt{2024ApJ...974..295D}; \citealt{2024ApJ...971L..20C}; \citealt{2026ApJ..1000L..48M}). 

\citet{2023ApJ...956..138S} investigated atmosphere model systematics in NICER pulse-profile analyses of PSR~J0030$+$0451 and PSR~J0740$+$6620, including fully ionized hydrogen and helium models. They found that none of the atmosphere cases significantly changed the inferred radius of PSR~J0740$+$6620, which they attributed potentially to the source's X-ray faintness, tighter external constraints, and/or view geometry. Our synthetic experiment is complementary to that result. We confirm their result at low source-to-background ratios, but demonstrate that the modeling assumption of atmospheric composition can introduce overwhelming systematic biases when analyzing higher-fidelity data.

In this paper, we check whether atmospheric composition mismatch can bias the inferred radius of a PSR~J0740$+$6620-like neutron star while remaining hidden inside a statistically acceptable fit. We generate synthetic NICER-like pulse-profile data assuming fully ionized hydrogen or fully ionized helium atmospheres, and fit the hydrogen-generated data with both atmosphere models and the helium-generated data with both models. In each case, we vary the ratio of hot spot counts to background counts while holding the total expected count number fixed at $\sim 5.5 \times 10^5$, which is the number observed from PSR~J0740$+$6620, in order to explore how the detectability of the composition mismatch depends on the relative strength of the pulsed signal. For each configuration, we perform input-parameter evaluations, full posterior sampling, goodness-of-fit assessments, and Bayesian evidence comparisons.

We find that in both directions of the atmospheric mismatch, the incorrect atmosphere model can yield statistically acceptable phase-channel and bolometric $\chi^2$ values after posterior exploration, yet still produce biased radius estimates. The effect is direction-dependent. Fitting hydrogen-generated data with a helium model can increase the inferred radius by more than $2\sigma$, whereas fitting helium-generated data with a hydrogen model produces smaller increases that remain within $2\sigma$ for the cases we examined. However, at the actual count level and spot-to-background ratio estimated for PSR~J0740$+$6620, fitting hydrogen-generated data with a helium model does not produce substantial bias in the inferred radius. Despite the statistical acceptability of these fits, the Bayesian evidence consistently favors the correct atmospheric model. The apparent asymmetry between the two mismatch directions should not be over-interpreted, as it may depend on the assumed geometry, count level, background conditions, and, in particular, we expect it to weaken for less compact stars.

The remainder of this paper is organized as follows. Section \ref{sec:atm} describes the effects of atmospheric beaming of the emergent radiation on the pulse profile. Section~\ref{sec:methods} describes our methods for generating and analyzing the synthetic data. Section~\ref{sec:results} presents our results. We discuss the implications of these results in Section~\ref{sec:discussion} and summarize our conclusions in Section~\ref{sec:conclusions}.

\begin{deluxetable*}{cccc}\caption{Primary Parameters in the Pulse Waveform Model}
\tablehead{
\colhead{Parameter} & \colhead{Definition} & \colhead{Assumed Prior}
& \colhead{Best-Fit Value}}
\startdata
$c^2R_e/(GM)$ & Inverse stellar compactness & $3.2-8.0$ & $4.52$\\
 $M$ & Gravitational mass & $\exp[-(M - 2.08 M_\odot)^{2}/2(0.09 M_\odot)^{2}]$ & $2.067$\\
 $\theta_{c,1}$ & Spot 1 inclination of center spot & $0.0-\pi$ rad & $0.834$\\
 $\Delta\theta_1$ & Spot 1 radius & $0.0 - 3.14$ rad & $0.087$\\
 $kT_{\rm eff,1}$ & Spot 1 effective temperature & $0.011-0.5$ keV & $0.098$\\
 $\Delta\phi_2$ & Spot 1 and 2 longitudinal offset & $0.0-1.0$ cycles & $0.577$\\
 $\theta_{c,2}$ & Spot 2 inclination of center spot & $0.0-\pi$ rad & $1.223$\\
 $\Delta\theta_2$ & Spot 2 radius & $0.0-3.14$ rad & $0.066$\\
 $kT_{\rm eff,2}$ & Spot 2 effective temperature & $0.011-0.5$ keV & $0.103$\\
 $\theta_{\rm obs}$ & Observer inclination & $1.44-1.62$ rad & $1.561$\\
 $N_H$ & Column density & $0.0-2.0 \times 10^{20}$ cm$^{-2}$ & $0.006$\\
 $D$ & Distance & $\exp[-(D - 1.136 \rm{kpc})^{2}/2(0.20 \rm{kpc})^{2}], D \geq 1.136 \rm{kpc}$ & $1.151$\\ 
  & & $\exp[-(D - 1.136 \rm{kpc})^{2}/2(0.18 \rm{kpc})^{2}], D \leq 1.136 \rm{kpc}$ & 
\enddata
\tablecomments{This table gives the pulse waveform model that \citet{2021ApJ...918L..28M} fit jointly to the NICER and XMM-Newton data on PSR~J0740$+$6620, the priors on the parameters, and the best-fit values.  \citet{2021ApJ...918L..28M} used flat priors of 0.0--3.0~rad for $\Delta\theta_1$ and $\Delta\theta_2$. In the present work, we adopted slightly wider priors of 0.0--3.14~rad for both parameters.  The 12 parameters listed here are the primary parameters in the pulse waveform model used in \citet{2021ApJ...918L..28M}. In the present work, we focus on the equatorial circumferential radius $R_e$.  We generated synthetic pulse waveform data by Poisson-sampling this best-fit waveform at the same 346.532~Hz rotational frequency assumed by \citet{2021ApJ...918L..28M}.  Unlike \citet{2021ApJ...918L..28M}, which analyzed synthetic joint NICER and XMM-Newton data, we generate and analyze only synthetic NICER data. The priors listed here are those used when fitting the model to our synthetic datasets.  See the text for further details.}
\label{table:parameters}
\end{deluxetable*}

\section{Atmospheric Beaming and its Effects on the Pulse Profile}\label{sec:atm}

PPM is less susceptible to atmospheric uncertainty than spectral fitting methods because the pulse waveform carries geometric information via phase-dependent flux modulation. The time-varying projection of a heated region (``hot spot") as the star rotates encodes the observer inclination, the spot colatitude, and the stellar compactness through the relativistic light-bending that maps surface emission angles to the observer's line of sight. Even if the beaming function is incorrect, the phase at which the spot appears and disappears and the overall depth of the modulation are primarily determined by the spot geometry and the star's compactness, reducing the freedom available for an incorrect atmosphere to absorb the mismatch. Indeed, analyses of synthetic data have shown that PPM radius estimates are robust against several other classes of systematic error. \citet{2013ApJ...776...19L} and \citet{2015ApJ...808...31M, 2016EPJA...52...63M} demonstrated that incorrect assumptions about the shapes and temperature distributions of the hot spots do not significantly bias the inferred radius, provided the fit is statistically acceptable. \citet{2025arXiv251116759H} showed that in joint NICER/XMM-Newton analyses, even a factor-of-five underestimate of the XMM-Newton background shifts the radius posterior by only $\sim 1\sigma$. These results have strengthened the case for PPM as a method for reliable radius inference. To our knowledge, however, no prior synthetic data study has systematically checked whether atmospheric composition mismatch, specifically, data generated with one atmosphere using a model that assumes another, can produce a comparably small or potentially larger bias.

In fully ionized atmospheres at the effective temperatures relevant to NICER ($kT_{\rm eff} \sim 0.1$~keV), the dominant sources of continuum opacity are free--free (inverse bremsstrahlung) absorption and electron scattering. The free--free absorption coefficient scales as $\alpha_{\rm ff} \propto Z^2 n_i n_e / \nu^3$ for ion charge $Z$, ion and electron number densities $n_i$ and $n_e$, and photon frequency $\nu$ (see, e.g., \citealt{1979rpa..book.....R}). The composition dependence of this expression is more subtle than the explicit $Z^2$ factor suggests. At a given mass density $\rho$, a fully ionized hydrogen atmosphere has $n_i = n_e = \rho/m_{\rm H}$, whereas a fully ionized helium atmosphere has $n_i = \rho/4m_{\rm H}$ and $n_e = \rho/2m_{\rm H}$. A helium nucleus is four times the mass of a proton and supplies only two electrons. As a result, helium offers fewer targets and fewer absorbers per gram. The reduction in $n_i$ cancels the factor $Z^2 = 4$ exactly, and the accompanying reduction in $n_e$ leaves $\alpha_{\rm ff}$ a factor of two lower for helium than for hydrogen at the same density and temperature.

\begin{figure}
    \centering
    \includegraphics[width=1.0\linewidth]{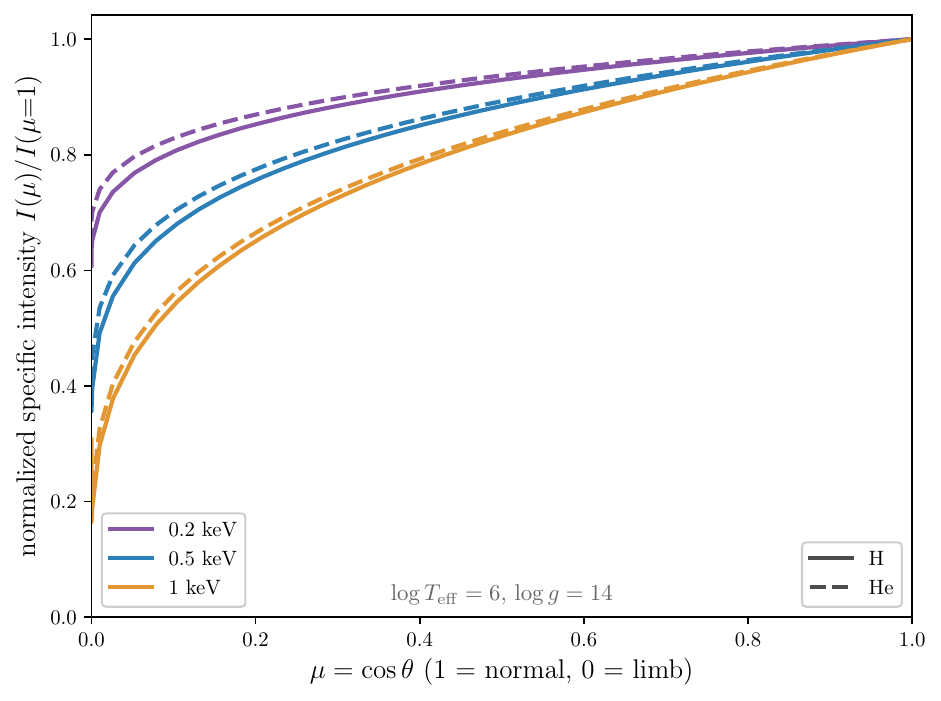}
    \caption{Emergent specific intensity as a function of emission angle (the beaming pattern) for hydrogen (solid) and helium (dashed) atmospheres at $\log T_{\rm{eff}} = 6$ and $\log g = 14$. The $\theta$ in $\mu = \cos\theta$ is measured from the local surface normal, so $\mu = 1$ is normal emission and $\mu = 0$ is the limb. Each curve is normalized to its emission value $I(\mu{=}1)$. Line colors denote photon energy (purple: $0.2$ keV; blue: $0.5$ keV; orange: $1$ keV). At every energy, the helium atmosphere retains more intensity towards grazing angles than hydrogen; its beaming is closer to isotropic, while hydrogen is more limb-darkened, and the limb darkening steepens with increasing photon energy.}
    \label{fig:beaming}
\end{figure}

The beaming pattern, i.e., the intensity as a function of the angle $\theta$ between the outgoing photon direction and the surface normal, is shaped by how the radiation decouples from the atmosphere, and the difference between the two compositions is therefore not a simple consequence of the ionic charge. On the basis of the scalings above, we expect it to be modest rather than dramatic. Figure~\ref{fig:beaming} shows the emergent beaming patterns for both of the model atmospheres. At every energy, the helium atmosphere retains more intensity toward grazing angles than hydrogen, meaning that its beaming is closer to isotropic, whereas hydrogen is more limb-darkened, consistent with previous calculations (see, e.g., \citealt{2021ApJ...914L..15B}; \citealt{2023ApJ...956..138S}). The two patterns are nonetheless similar in overall shape, and the difference between them is small. Its magnitude depends on the effective temperature, surface gravity, and photon energy.

\begin{figure}
    \centering
    \includegraphics[width=1.0\linewidth]{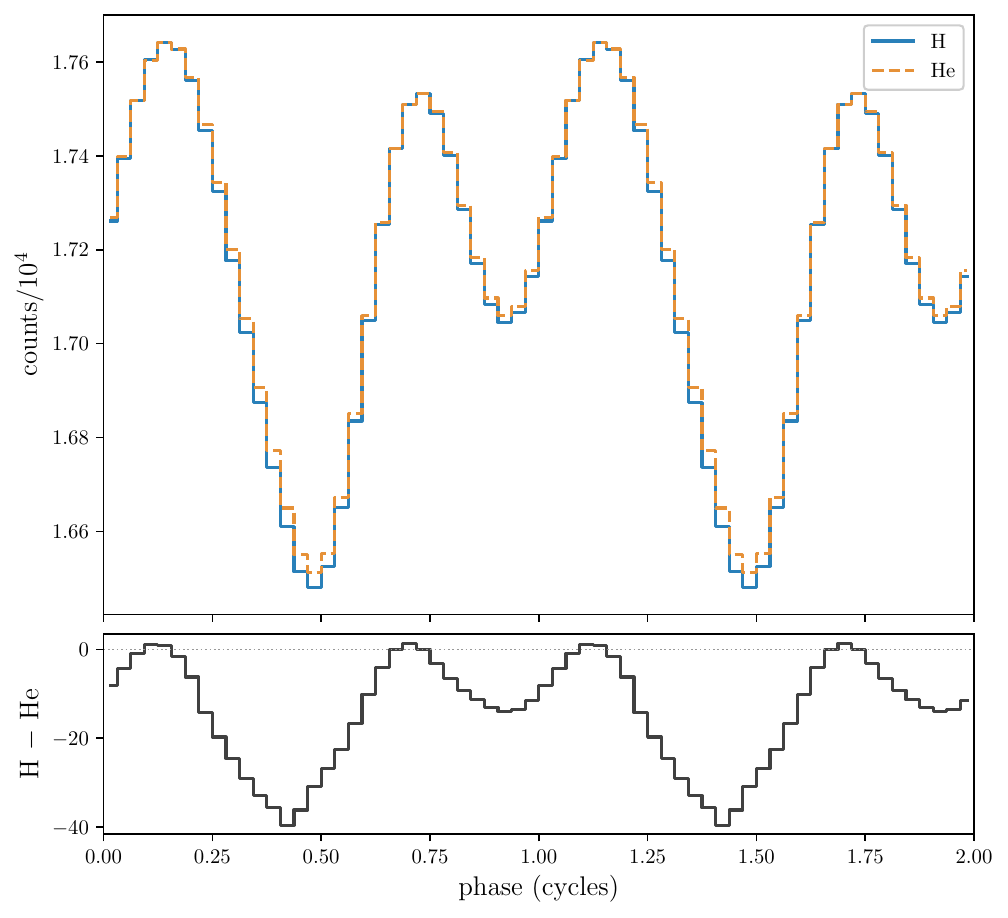}
    \caption{Synthetic pulse profile for a PSR~J0740$+$6620-like geometry with hydrogen (blue, solid) and helium (orange, dashed) atmospheres at the realistic spot-to-background ratio. The lower panel shows $\rm{H} - \rm{He}$. The pulsed signal is a few percent of the total counts and the two compositions are statistically indistinguishable, with fractional modulations $f_{\rm{H}} = 0.034$ and $f_{\rm{He}} = 0.033$.}
    \label{fig:pulse-1x}
\end{figure}

\begin{figure}
    \centering
    \includegraphics[width=1.0\linewidth]{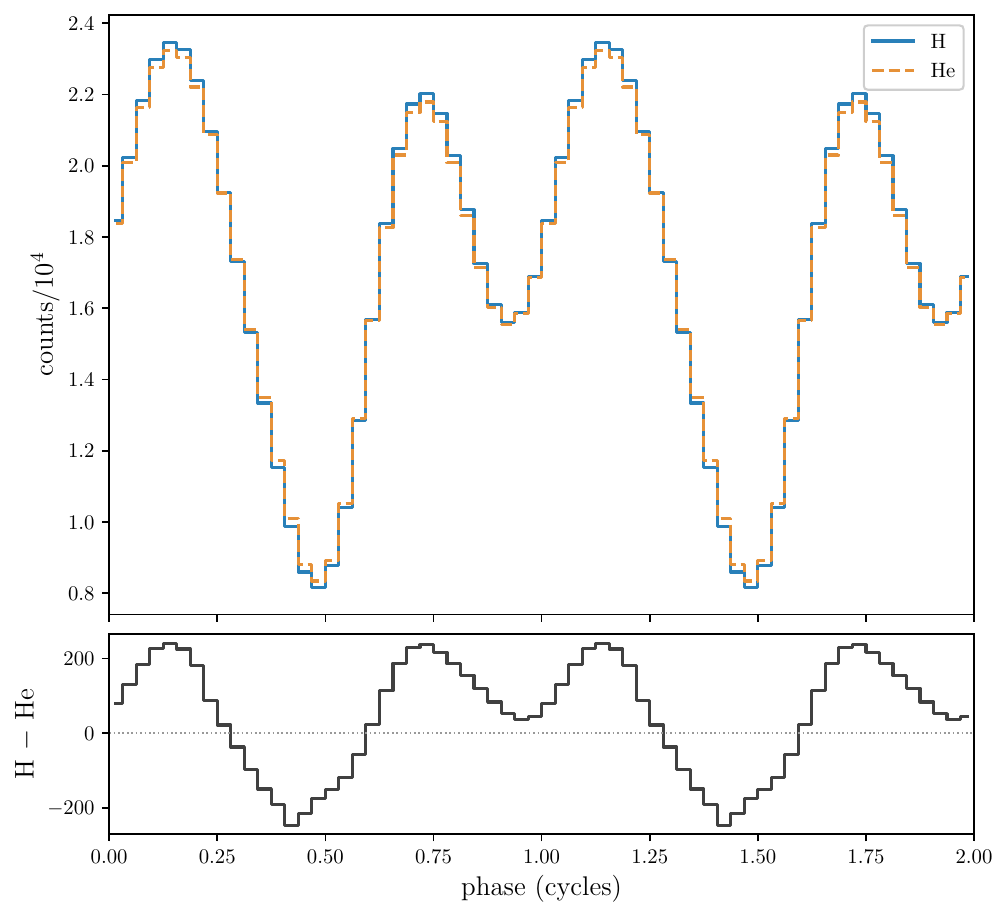}
    \caption{Synthetic pulse profile for a PSR~J0740$+$6620-like geometry with the spot contribution amplified by a factor of $60$. The model assuming a hydrogen atmosphere produces a marginally deeper pulse with fractional modulations $f_{\rm{H}} = 0.484$ and $f_{\rm{He}} = 0.471$, which is consistent with limb darkening for a hydrogen atmosphere.}
    \label{fig:pulse-60x}
\end{figure}

For PPM, the consequence of this beaming difference is that a helium atmosphere directs relatively more radiation toward grazing angles compared to a hydrogen atmosphere. When the star rotates and the hot spot sweeps across the observer's line of sight, the angular distribution of the emitted radiation determines how steeply the observed flux rises and falls with pulse phase. A more limb-darkened (hydrogen-like) pattern produces a sharper pulse profile with deeper modulation for a given geometry, whereas a more isotropic (helium-like) pattern produces a broader, shallower pulse. We illustrate this in Figures \ref{fig:pulse-1x} and \ref{fig:pulse-60x}, which compare synthetic pulse profiles generated with hydrogen and helium atmospheres for PSR~J0740$+$6620-like geometry. At the spot-to-background ratio of PSR~J0740$+$6620, the two compositions yield nearly identical waveforms, with fractional modulations of $f_{\rm{H}} = 0.034$ and $f_{\rm{He}} = 0.033$ (Figure \ref{fig:pulse-1x}). Amplifying the ratio by a factor of 60 to suppress the background exposes the shape difference, yielding fractional modulations, $f_{\rm{H}} = 0.484$ and $f_{\rm{He}} = 0.471$ (Figure \ref{fig:pulse-60x}). In both regimes, the hydrogen atmosphere produces the slightly deeper, sharper pulse expected from its stronger limb darkening.

This is particularly relevant for PSR~J0740$+$6620, a $\sim 2.1~M_\odot$ pulsar (\citealt{2021ApJ...915L..12F}) whose high mass makes it valuable for constraining the EOS of cold, dense matter at densities above those accessible through observations of $\sim 1.4~M_\odot$ neutron stars (see, e.g., \citealt{2021ApJ...918L..28M}; \citealt{2021ApJ...918L..27R}; \citealt{2024ApJ...974..294S}; \citealt{2024ApJ...974..295D}). The NICER observations of PSR~J0740+6620 are severely background-dominated \citep{2021ApJ...918L..26W}, with the estimated ratio of hot spot counts to total counts in the NICER band being of order a few percent. This means that the pulsed signal from the spots is superimposed on a large background of counts from other X-ray sources in the field. This low spot-to-background ratio has two competing implications for the composition mismatch problem. It may make the waveform less sensitive to the beaming pattern, because the beaming-dependent modulation constitutes a smaller fraction of the total signal and the statistical uncertainties in the pulse shape are large. On the other hand, if the beaming mismatch does shift the inferred parameters, the large background means that the shift may be harder to detect through standard goodness-of-fit diagnostics.

\section{Methods}\label{sec:methods}

In this section we describe our procedure for generating synthetic NICER-like data for fully ionized hydrogen and helium atmospheres, and for fitting each dataset with both atmosphere models.  We first describe the baseline configuration (\S\ref{subsec:baseline}) and then the spot-to-background scaling experiment (\S\ref{subsec:spot_bg}), the synthetic data generation procedure (\S\ref{subsec:datagen}), the likelihood and normalization treatment (\S\ref{subsec:likelihood}), the two-stage analysis (\S\ref{subsec:analysis}), the fit diagnostics (\S\ref{subsec:diagnostics}), the Bayesian evidence calculation (\S\ref{subsec:evidence}), and the radius bias diagnostic (\S\ref{subsec:radius_bias}).

\subsection{Baseline J0740-like Configuration}\label{subsec:baseline}

We adopt the two-uniform-temperature spot geometry and best-fit 12-parameter solution from \citet{2021ApJ...918L..28M} as our reference configuration.  The parameters are the gravitational mass $M$, equatorial circumferential radius $R_e$, observer inclination $i$, hydrogen column density $N_H$, source distance $d$, and the colatitudes, angular radii, and effective temperatures of the two hot spots. Table~\ref{table:parameters} lists these 12 parameters, the ranges of the priors \citet{2021ApJ...918L..28M} used when they fit this pulse waveform model to the actual NICER and XMM-Newton data, and the best-fit values of these parameters. As noted in Table~\ref{table:parameters}, we adopted slightly wider priors for the spot angular radii $\Delta\theta_1$ and $\Delta\theta_2$ than were used by \citet{2021ApJ...918L..28M}. The purpose of choosing a specific reference configuration is not to exhaust all possible pulsar geometries, but to check the consequences of atmospheric composition mismatch in a realistic setting.

We set the total expected number of counts in the synthetic datasets to $N_{\rm counts} \sim 5.5 \times 10^5$, consistent with the NICER data on  PSR~J0740$+$6620 used in \citet{2021ApJ...918L..28M}.  The baseline ratio of hot-spot counts to background counts, $R_0 \equiv S_0/B_0$, reflects the background-dominated conditions of the actual NICER observations, in which only a few percent of the total counts originate from the pulsed hot-spot emission.

\subsection{Spot-to-Background Ratio Experiment}\label{subsec:spot_bg}

To explore how the consequences of atmospheric composition mismatch depend on the relative strength of the pulsed signal, we systematically vary the ratio of hot-spot counts to background counts while holding the total expected count number fixed. We define the multiplier $m$ as
\begin{equation}\label{eq:target_ratio}
    m\equiv\frac{R_{\rm synth}}{R_0},
\end{equation}
where $R_{\rm synth}$ is the rescaled spot-to-background count ratio and $R_0\equiv\frac{S_0}{B_0}$ is the baseline ratio.  The total number of counts is:
\begin{equation}\label{eq:total_counts}
    N_{\rm counts}=S_{\rm synth} + B_{\rm synth},
\end{equation}
where $S_{\rm synth}$ and $B_{\rm synth}$ are the rescaled total spot and background counts, respectively.  We require their ratio to satisfy
\begin{equation}\label{eq:new_ratio}
    \frac{S_{\rm synth}}{B_{\rm synth}} = R_{\rm synth}\,.
\end{equation}

For clarity, the count normalization used in both generation and likelihood evaluations can be written explicitly. For each multiplier we define
\begin{equation}\label{eq:scale_factors}
                a_m \equiv \frac{S_{\rm synth}}{S_0}, \qquad
                b_m \equiv \frac{B_{\rm synth}}{B_0}.
\end{equation}
If $S^{(0)}_{jk}$ and $B^{(0)}_{jk}$ are the baseline expected spot and background counts in energy channel $j$ and phase bin $k$, the mean counts used to generate the synthetic data are
\begin{equation}\label{eq:slambda}
            \lambda^{(m)}_{jk}=a_m S^{(0)}_{jk}+b_m B^{(0)}_{jk}.
\end{equation}
The waveform calculation implements the spot scaling through the exposure time: for each synthetic dataset we use $T_{\rm obs}^{(m)}=a_m T_{\rm obs}^{(0)}$. Because the predicted spot counts scale linearly with exposure, evaluation of the same 12 physical input parameters for that dataset predicts $a_m S_0=S_{\rm synth}$ spot counts, rather than the original $S_0$. Thus, the increased spot normalization is not supplied by an additional signal normalization parameter. The factor $b_m$ is used to construct the synthetic background. The treatment of the background during fitting is described in \S\ref{subsec:likelihood}.

By choosing $m$ values  1, 20, 40, 60, 80, and 100, we explore configurations ranging from the realistic, heavily background-dominated regime of PSR~J0740$+$6620 to strongly spot-dominated scenarios similar to the millisecond X-ray pulsar PSR~J0437-4715. The high multiplier cases are controlled experiments and are not intended to represent PSR~J0740$+$6620 itself with otherwise unchanged observational properties. A source with such a large pulsed fraction would generally require the observational setup and background assumptions to be reconsidered. We also explored higher multipliers such as 120, 140, 160 and 180, but these multipliers do not add significant information, given the already very high signal-to-noise.

\subsection{Synthetic Data Generation}\label{subsec:datagen}

The hot-spot emission spectra and beaming patterns are computed using the NSX code for fully ionized model atmospheres \citep{2001MNRAS.327.1081H}, following the same approach adopted in \citet{2021ApJ...918L..28M}.  We note that the surface gravity of a neutron star is large enough that the lightest element present is expected to settle to the top of the atmosphere within seconds to minutes \citep[based on extrapolations of calculations in][]{1980ApJ...235..534A}.  As hydrogen is the most abundant element in the universe, and PSR~J0740$+$6620 likely underwent prolonged accretion to reach its current spin frequency, a pure hydrogen atmosphere is a well-motivated assumption \citep{1987ApJ...313..718R,2019ApJ...887L..26B}.  However, the true surface composition is uncertain. A helium atmosphere is possible if the accreted material was hydrogen-poor or if subsequent processes altered the composition.  We therefore check both fully ionized hydrogen and helium atmospheres.

The NICER data for PSR~J0740$+$6620 are recorded in 94 energy channels and are binned into 32 rotational phase bins. We refer to each of the $32\times94 = 3008$ combinations of a phase bin and an energy channel as a phase-channel bin. For the synthetic hydrogen data, we compute the expected phase-channel count distribution using the fully ionized hydrogen atmosphere model. The baseline per-bin spot and background distributions are scaled to the new totals $S_{\rm synth}$ and $B_{\rm synth}$ from \S\ref{subsec:spot_bg} without changing their shape, and we draw a Poisson realization of the counts in each phase-channel bin.  For each multiplier $m$ we perform an independent Poisson realization.  The resulting dataset is then fit with both the hydrogen atmosphere model and the helium atmosphere model. For the synthetic helium data, we repeat the procedure using the fully ionized helium atmosphere model to generate the expected counts.  Each multiplier again corresponds to an independent Poisson realization, and the resulting dataset is fit with both the helium atmosphere model and the hydrogen atmosphere model. 

In both cases, the synthetic data are generated using the reference 12-parameter configuration described in \S\ref{subsec:baseline}.  The only difference between the two families is the atmosphere model.

\subsection{Likelihood and normalization}\label{subsec:likelihood}

We evaluate the NICER data using the phase-and energy-resolved Poisson likelihood adopted in previous analyses of PSR~J0740$+$6620 \citep{2021ApJ...918L..28M,2024ApJ...974..295D}. For a model predicting $m_{jk}$ counts in energy channel $j$ and phase bin $k$, with $d_{jk}$ observed counts, the log likelihood is

\begin{equation}
    \ln{\mathcal{L}}=\sum_{j,k}\left[d_{jk}\ln m_{jk}-m_{jk}\right],
\end{equation}
where terms independent of the model have been omitted. As in \citet{2024ApJ...974..295D}, a phase-independent background contribution is allowed independently in each NICER energy channel and is marginalized over, together with the overall rotational phase.

The spot normalization for each scaled synthetic dataset is fixed by the dataset-specific exposure described in Section~\ref{subsec:spot_bg}; there is no additional free signal-normalization parameter. The background normalization is handled separately through the NICER background marginalization described above.

\subsection{Two-Stage Analysis Procedure}\label{subsec:analysis}

In the first stage, we evaluate the likelihood with the 12 primary parameters held fixed at their input values (`Best-fit value' from Table \ref{table:parameters}). The 94 energy-channel background and the overall phase shift of the waveform are marginalized over analytically. We refer to this throughout as the input-parameter evaluation. It provides an initial diagnostic of how much the incorrect atmosphere degrades the fit before the primary parameters are allowed to adjust.

In the second stage, we sample over all 12 primary parameters by performing a full Bayesian posterior exploration using \texttt{pocoMC} \citep[preconditioned Monte Carlo;][]{2022MNRAS.516.1644K,2022JOSS....7.4634K}, allowing all model parameters to vary within their priors. The sampler stochastically explores parameter space for each atmosphere assumption.  From the resulting \texttt{pocoMC} run, we extract the posterior samples and Bayesian evidence. For the goodness-of-fit diagnostics reported below, we evaluate $\chi^2$ at the maximum-likelihood sample, defined as the stored \texttt{pocoMC} sample with the largest $\ln\mathcal{L}$.

We chose \texttt{pocoMC} over the nested sampler+Markov Chain Monte Carlo hybrid strategy employed in \citet{2019ApJ...887L..24M,2021ApJ...918L..28M,2024ApJ...974..295D} and \citet{2025arXiv251116759H} for several reasons.  First, \texttt{pocoMC} is designed to handle the complex, multimodal posterior surfaces that arise in pulse-profile modeling, using a normalizing-flow-based preconditioner to accelerate convergence in high-dimensional parameter spaces.  Second, \texttt{pocoMC} provides a direct estimate of the Bayesian evidence as a byproduct of the sampling, eliminating the need for a separate nested sampling run.  Third, previous work has shown that some popular nested samplers, such as MultiNest, can underestimate posterior uncertainties when the number of live points is insufficient or the sampling efficiency parameter is too large \citep{2024ApJ...974..295D, 2024ApJ...974..294S, 2025PhRvD.112b3008H}, requiring subsequent MCMC continuation to broaden the credible intervals to their converged values.  The \texttt{pocoMC} sampler avoids this two-step procedure by producing converged posteriors directly and returning reliable evidence estimates.  For this study, in which we analyze 24 independent configurations (4 test cases $\times$ 6 multipliers), the computational efficiency of a single-stage sampler is a practical advantage.

\subsection{Fit Diagnostics}\label{subsec:diagnostics}

We assess the quality of each fit using two $\chi^2$ statistics. The phase-channel $\chi^2$ uses the standard Pearson $\chi^2$ computed over all phase-channel bins, i.e.,
\begin{equation}\label{eq:chi2_pc}
    \chi^2 = \sum_{i} \frac{(d_i - m_i)^2}{m_i}\,,
\end{equation}
where $d_i$ and $m_i$ are the observed and model-predicted counts in bin $i$, and the sum runs over all $3008$ phase-channel bins.  This statistic is sensitive to localized discrepancies in individual bins.

The bolometric $\chi^2$ is computed by first summing the counts over energy channels to produce a phase-only pulse profile, and then computing the $\chi^2$ of this collapsed profile.  The bolometric $\chi^2$ checks whether the overall pulse shape is well reproduced when integrated over the photon energy.

Next, we compared the resulting value of $\chi^2$ with the expected $\chi^2$ distribution for the appropriate number of degrees of freedom in the input-parameter and sampled analyses, respectively.  We report the probability of obtaining a $\chi^2$ value at least as large as the observed one under the null hypothesis that the model is correct (i.e., the $p$-value).  We note that acceptable $\chi^2$ values may not guarantee that the assumed atmospheric composition is correct in all cases. A model that passes these checks can still return a biased radius posterior, as we demonstrate in \S\ref{sec:results}. Nevertheless, $\chi^2$ remains a valuable first diagnostic. Poor $\chi^2$ values at the input-parameter stage correctly identify composition mismatch at high spot-to-background ratios before posterior exploration (see Tables~\ref{tab:fixed_H_data} and~\ref{tab:fixed_He_data}). Acceptable $\chi^2$ values after sampling confirm that a plausible fit has been found using the wrong atmosphere, and that this fit is worth scrutinizing further.

For the phase-channel $\chi^2$, the number of degrees of freedom is determined by the number of phase-channel bins minus the number of model parameters that are free to vary.  In the input-parameter evaluation, all 12 primary parameters are held fixed, but the 94 energy-channel background levels and one overall rotational phase are marginalized analytically, giving $3008 - 94 - 1 = 2913$ degrees of freedom.  In the fully sampled analysis, all 12 primary parameters are also free, giving $3008 - 94 - 1 - 5 = 2908$ degrees of freedom, where the effective number of waveform parameters influencing the fit is approximately 5 (determined from synthetic data checks following the procedure described in \citealt{2024ApJ...974..295D}).  For the bolometric $\chi^2$, we sum the counts over energy channels to produce 32 phase bins and adopt an effective number of parameters of approximately 5 plus the overall phase and a single background term, giving approximately $32 - 5 - 2 = 25$ degrees of freedom in the sampled case and $32 - 2 = 30$ in the input-parameter case.

\subsection{Bayesian Evidence}\label{subsec:evidence}

To compare the hydrogen and helium atmosphere models for each synthetic dataset, we compute the Bayesian evidence $\mathcal{Z}$ for each model.  The evidence is defined as the integral of the likelihood over the prior, i.e.,
\begin{equation}\label{eq:evidence}
    \mathcal{Z} = \int \mathcal{L}(\theta)\,\pi(\theta)\,d\theta\,,
\end{equation}
where $\mathcal{L}(\theta)$ is the likelihood, $\pi(\theta)$ is the prior, and $\theta$ denotes the vector of model parameters.  We report the log evidence ratio
\begin{equation}\label{eq:delta_logZ}
    \Delta \ln \mathcal{Z} = \ln\mathcal{Z}_{\rm correct} - \ln \mathcal{Z}_{\rm incorrect}\,,
\end{equation}
where ``correct'' and ``incorrect'' models refer to the atmospheric composition used to generate the synthetic data and the mismatched atmospheric model, respectively. Thus, $\Delta\ln\mathcal{Z} > 0$ favors the correct atmosphere model, while $\Delta\ln\mathcal{Z} < 0$ favors the incorrect model.

We do not adopt fixed qualitative thresholds for interpreting the magnitude of $\Delta\ln\mathcal{Z}$. Such thresholds are arbitrary and do not account for uncertainty in the evidence estimates or for variation introduced by the independent Poisson realizations used for the synthetic datasets. We therefore report the evidence differences and interpret them comparatively across the models and synthetic experiments.

Unlike the $\chi^2$ statistic, which evaluates the quality of the fit at the best-fit point, the evidence integrates over the entire posterior volume, which makes the evidence sensitive to model adequacy. A model that achieves an acceptable fit but requires a fine-tuned region of parameter space will be penalized relative to a model that fits comparably well over a broader parameter range.  We use the evidence to check whether the correct atmospheric model can be identified even when both models yield statistically acceptable $\chi^2$ values.

\subsection{Radius Bias Diagnostic}\label{subsec:radius_bias}
 
To quantify the bias in the inferred radius introduced by assuming the incorrect atmospheric composition, we compute the percentile location of the injected radius $R_{\rm inj}$ within the radius posterior obtained for the incorrect atmosphere fit.  We adopt the convention that a percentile below the ${\sim}\,2.3$rd or above the ${\sim}\,97.7$th places $R_{\rm inj}$ outside the $2\sigma$ credible interval of the posterior, which indicates a potentially worrisome bias.  Intermediate percentiles indicate a shift that, while potentially concerning, remains within $2\sigma$ and does not constitute a definitive detection of significant bias given the statistical uncertainties inherent in a single Poisson realization.

\section{Results}\label{sec:results}

\begin{figure*}[!htbp]
\centering
\includegraphics[width=\textwidth]{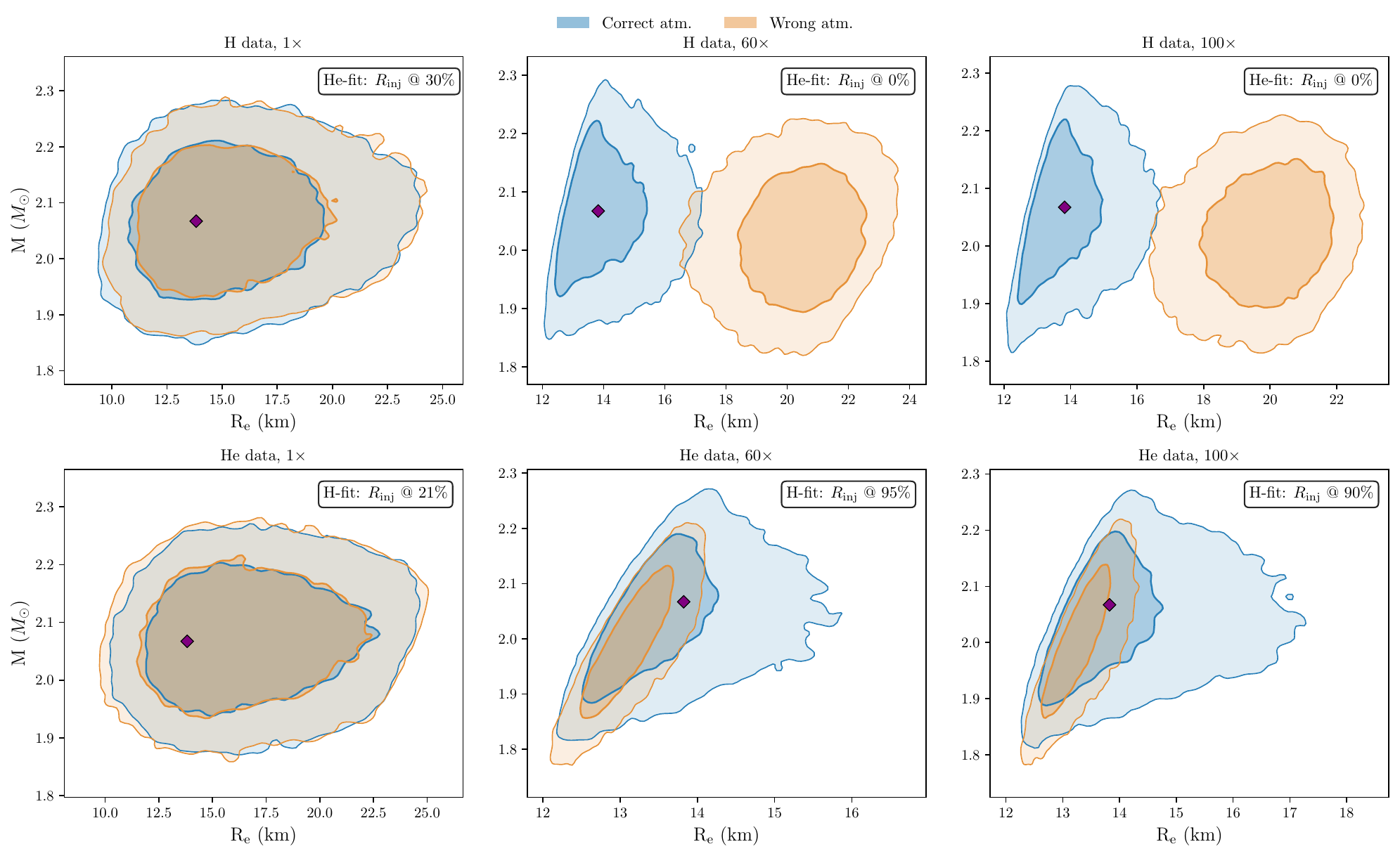}
\caption{Mass--radius posterior distributions for selected multipliers ($1\times$, $60\times$, $100\times$).  Blue contours show the correct atmosphere fit ($1\sigma$ and $2\sigma$ credible regions), and orange contours show the incorrect atmosphere fit.  The purple diamond marks the injected values ($R_{\rm inj} = 13.82$~km, $M_{\rm inj} = 2.067\;M_\odot$).  \textit{Top row}: hydrogen-generated data, showing the correct H~fit and incorrect He~fit.  The percentile of $R_{\rm inj}$ in the correct H-fit posterior is 33rd, 51st, and 58th at $1\times$, $60\times$, and $100\times$, respectively, confirming that the correct model recovers the injected radius without significant bias.  The corresponding incorrect He-fit percentiles are 30th, 0.06th, and 0.07th, showing that the He-fit posterior is displaced to substantially larger radii at $60\times$ and $100\times$.  \textit{Bottom row}: helium-generated data, showing the correct He~fit and incorrect H~fit.  The correct He-fit percentiles are 18th, 71st, and 58th, while the incorrect H-fit percentiles are 21st, 95th, and 90th.  The H-fit posterior remains closer to the truth for all three multipliers.  The percentile of $R_{\rm inj}$ in the incorrect model posterior is annotated in each panel.}
\label{fig:showcase_MR}
\end{figure*}

\begin{figure*}
\centering
\includegraphics[width=\textwidth]{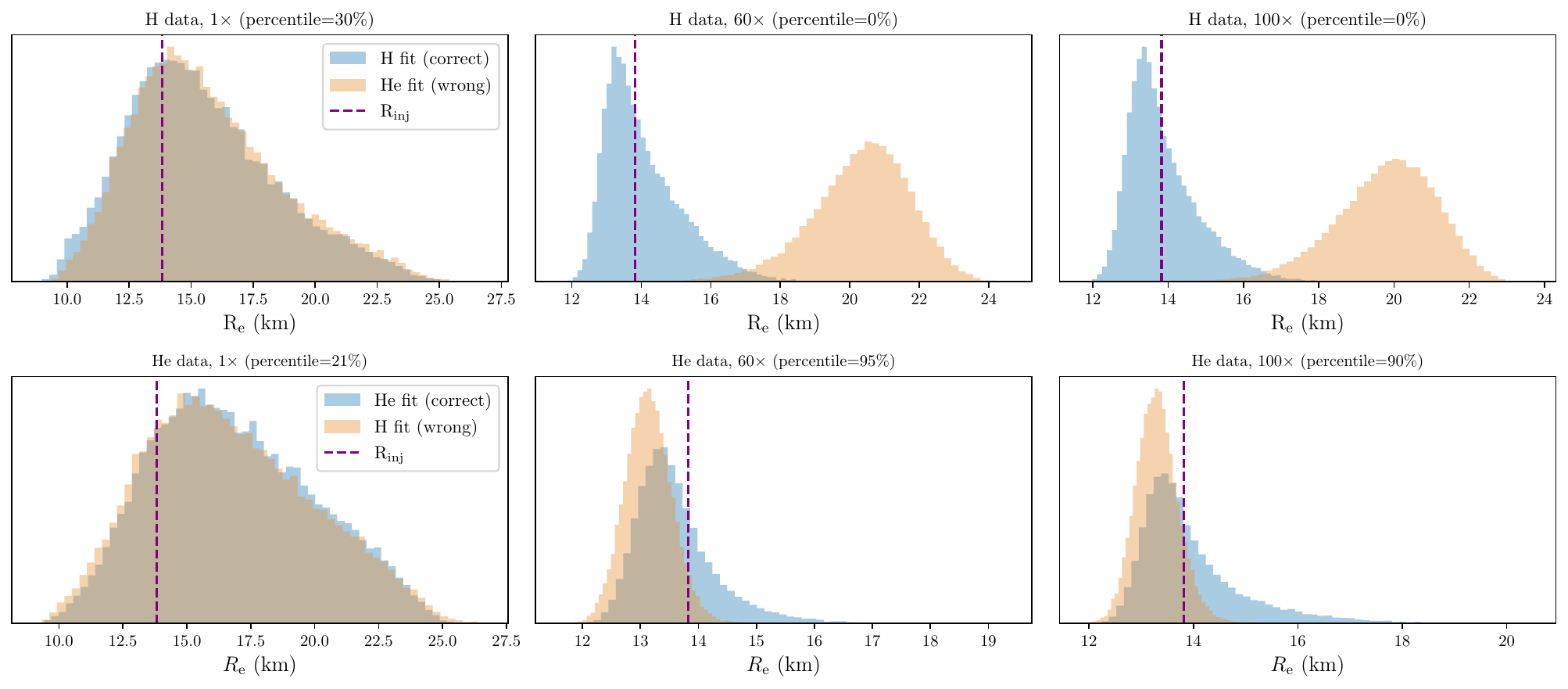}
\caption{One-dimensional radius posteriors for selected multipliers ($1\times$, $60\times$, $100\times$).  The blue histograms show the correct atmosphere fit, and the orange histograms show the incorrect atmosphere fit.  The dashed purple line marks the injected radius.  \textit{Top row}: hydrogen-generated data.  The percentile of $R_{\rm inj}$ in the correct H-fit posterior is 33rd, 51st, and 58th, whereas in the incorrect He-fit posterior it is 30th, 0.06th, and 0.07th.  At $60\times$ and $100\times$, the He-fit (incorrect) posterior is shifted well above $R_{\rm inj}$.  \textit{Bottom row}: helium-generated data.  The correct He-fit percentiles are 18th, 71st, and 58th, and the incorrect H-fit percentiles are 21st, 95th, and 90th.  The bias is smaller when incorrectly fitting H to He data than when incorrectly fitting He to H data.}
\label{fig:1D_radius}
\end{figure*}

We present our results in three layers. We first describe the input-parameter diagnostics (\S\ref{subsec:fixed_param}), then the full posterior sampling results (\S\ref{subsec:sampled_fits}--\S\ref{subsec:He_to_H}), and finally the radius posterior and Bayesian evidence behavior in both mismatch directions (\S\ref{subsec:comparison}--\S\ref{subsec:evidence_results}).

\subsection{Input-parameter Incorrect Atmosphere Checks}\label{subsec:fixed_param}

Before allowing the model parameters to vary, we evaluate the likelihood at the 12-parameter solution under both the correct and incorrect atmosphere assumptions for each spot-to-background ratio multiplier.  Tables~\ref{tab:fixed_H_data} and \ref{tab:fixed_He_data} summarize these input-parameter diagnostics.

\begin{deluxetable}{clc cccc}
\tabletypesize{\footnotesize}
\setlength{\tabcolsep}{2.0pt}
\caption{Input-parameter evaluation for hydrogen-generated synthetic data\label{tab:fixed_H_data}}
\tablehead{
  & & & \multicolumn{2}{c}{H atm.\ (correct)} & \multicolumn{2}{c}{He atm.\ (incorrect)} \\[-4pt]
  \colhead{Scale} & \colhead{$R_{\rm synth}$} & \colhead{$f_{\rm pulse}$}
  & \colhead{$\chi^2_{\rm pc}$ (Prob.)}
  & \colhead{$\chi^2_{\rm b}$ (Prob.)}
  & \colhead{$\chi^2_{\rm pc}$ (Prob.)}
  & \colhead{$\chi^2_{\rm b}$ (Prob.)}
}
\startdata
$1\times$   & 0.064 & 0.034 & 2878 (.673) & 32.3 (.354) & 2877 (.679) & 31.5 (.390) \\
$20\times$  & 1.282 & 0.334 & 2824 (.878) & 31.1 (.410) & 2943 (.345) & 48.5 (.018) \\
$40\times$  & 2.565 & 0.435 & 2913 (.494) & 22.3 (.843) & 3141 (.002) & 60.8 (.001) \\
$60\times$  & 3.847 & 0.484 & 3025 (.072) & 47.2 (.024) & 3178 (.000) & 64.2 (.000) \\
$80\times$  & 5.130 & 0.513 & 2817 (.896) & 19.1 (.938) & 3097 (.009) & 75.7 (.000) \\
$100\times$ & 6.412 & 0.532 & 2996 (.139) & 41.0 (.086) & 3242 (.000) & 114 (.000) \\
\enddata
\tablecomments{Input-parameter $\chi^2$ values and probabilities at the reference 12-parameter solution.  ``Scale'' is the spot-to-background ratio multiplier $m$ (see \S\ref{subsec:spot_bg}). $R_{\rm synth}$ is the corresponding target spot-to-background ratio. $f_{\rm pulse}$ is the fractional modulation of the expected bolometric synthetic profile, $f_{\rm pulse}\equiv(C_{\max}-C_{\min})/(C_{\max}+C_{\min})$, and is included to show pulse strength with scaled configuration.  $\chi^2_{\rm pc}$ is the phase-channel $\chi^2$ and $\chi^2_{\rm b}$ is the bolometric $\chi^2$. The values in parentheses are $p$-values (the probability of obtaining a $\chi^2$ at least this large if the model is correct).  Degrees of freedom: 2913 (phase-channel) and 30 (bolometric).  The hydrogen model yields acceptable probabilities at all multipliers.  The helium model is strongly disfavored for multipliers $\geq 40\times$.}
\end{deluxetable}

\begin{deluxetable}{clc cccc}
\tabletypesize{\footnotesize}
\setlength{\tabcolsep}{2.0pt}
\caption{Input-parameter evaluation for helium-generated synthetic data\label{tab:fixed_He_data}}
\tablehead{
  & & \multicolumn{2}{c}{He atm.\ (correct)} & \multicolumn{2}{c}{H atm.\ (incorrect)} \\[-4pt]
  \colhead{Scale} & \colhead{$R_{\rm synth}$} & \colhead{$f_{\rm pulse}$}
  & \colhead{$\chi^2_{\rm pc}$ (Prob.)}
  & \colhead{$\chi^2_{\rm b}$ (Prob.)}
  & \colhead{$\chi^2_{\rm pc}$ (Prob.)}
  & \colhead{$\chi^2_{\rm b}$ (Prob.)}
}
\startdata
$1\times$   & 0.061 & 0.033 & 2950 (.312) & 17.2 (.970) & 2951 (.308) & 18.4 (.952) \\
$20\times$  & 1.224 & 0.325 & 2890 (.617) & 25.6 (.696) & 2947 (.326) & 41.4 (.080) \\
$40\times$  & 2.447 & 0.423 & 2910 (.513) & 30.5 (.438) & 3049 (.039) & 65.6 (.000) \\
$60\times$  & 3.671 & 0.471 & 2996 (.138) & 36.5 (.194) & 3196 (.000) & 77.8 (.000) \\
$80\times$  & 4.894 & 0.499 & 2896 (.583) & 26.9 (.630) & 3123 (.004) & 87.9 (.000) \\
$100\times$ & 6.118 & 0.517 & 2906 (.535) & 25.1 (.721) & 3103 (.007) & 77.0 (.000) \\
\enddata
\tablecomments{As in Table~\ref{tab:fixed_H_data}, but for helium-generated synthetic data.  The hydrogen model becomes strongly disfavored for multipliers $\geq 40\times$. }
\end{deluxetable}

For the hydrogen-generated synthetic data (Table~\ref{tab:fixed_H_data}), the helium atmosphere evaluations produce increasingly poor phase-channel and bolometric $\chi^2$ values at multipliers of $40\times$ and above. The bolometric $\chi^2$ rises from $\sim 31$ at $1\times$ to $114$ at $100\times$, while the correct hydrogen model maintains bolometric $\chi^2$ values in the range $\sim 19$--$47$.  The phase-channel $\chi^2$ shows a similar pattern, with the incorrect atmosphere values climbing above $3100$ while the correct atmosphere values remain near the expected value for 2913 degrees of freedom.

For the helium-generated synthetic data (Table~\ref{tab:fixed_He_data}), the pattern is qualitatively similar. The hydrogen atmosphere evaluations produce elevated phase-channel and bolometric $\chi^2$ values at multipliers of $60\times$ and above.  The bolometric $\chi^2$ under the incorrect hydrogen assumption reaches values up to $\sim 88$, compared with $\sim 17$--$36$ for the correct helium model.

These results show that at the fixed reference configuration, the incorrect atmosphere can appear strongly disfavored, especially at higher spot-to-background ratios where the pulsed signal contributes a larger fraction of the total counts.  However, as we show in the following subsections, these initial diagnostics do not predict the outcome of full posterior exploration.

\subsection{Full Posterior Sampling Recovers Acceptable Fits}\label{subsec:sampled_fits}

After full posterior exploration with \texttt{pocoMC}, the incorrect atmosphere model recovers statistically acceptable phase-channel and bolometric $\chi^2$ values for all checked spot-to-background ratios, in both mismatch directions.  Despite the poor input-parameter evaluations documented in \S\ref{subsec:fixed_param}, the sampler identifies parameter combinations that substantially improve the fit to the incorrect model.
 
\begin{deluxetable}{c cccc c}
\tabletypesize{\footnotesize}
\setlength{\tabcolsep}{2pt}
\caption{Maximum-likelihood diagnostics for hydrogen-generated synthetic data\label{tab:pocomc_H_data}}
\tablehead{
  & \multicolumn{2}{c}{H atm.\ (correct)} & \multicolumn{2}{c}{He atm.\ (incorrect)} \\[-4pt]
  \colhead{$R_{\rm synth}$}
  & \colhead{$\chi^2_{\rm pc}$ (Prob.)}
  & \colhead{$\chi^2_{\rm b}$ (Prob.)}
  & \colhead{$\chi^2_{\rm pc}$ (Prob.)}
  & \colhead{$\chi^2_{\rm b}$ (Prob.)}
  & \colhead{$\Delta\!\ln\!\mathcal{Z}$}
}
\startdata
0.064  & 2876 (.659) & 30.6 (.203) & 2875 (.664) & 30.6 (.202) & $-0.9$ \\
1.282  & 2812 (.898) & 28.5 (.287) & 2829 (.851) & 30.7 (.199) & $+10.8$ \\
2.565  & 2900 (.536) & 20.4 (.727) & 2924 (.415) & 20.5 (.721) & $+14.2$ \\
3.847  & 3011 (.089) & 36.2 (.068) & 3025 (.064) & 42.7 (.015) & $+9.7$ \\
5.130  & 2812 (.898) & 18.9 (.803) & 2852 (.769) & 24.1 (.516) & $+23.1$ \\
6.412  & 2983 (.162) & 32.7 (.138) & 2988 (.147) & 37.7 (.050) & $+4.1$ \\
\enddata
\tablecomments{$\chi^2$ values and probabilities evaluated at the maximum-likelihood \texttt{pocoMC} sample for the hydrogen-generated data.  Degrees of freedom: 2908 ($\chi^2_{\rm pc}$) and 25 ($\chi^2_{\rm b}$). For $\Delta\ln\mathcal{Z} = \ln\mathcal{Z}_{\rm H} - \ln\mathcal{Z}_{\rm He}$, the positive values favor the hydrogen model.  Both models yield acceptable $\chi^2$ values at every multiplier, yet the evidence consistently identifies the correct model in all spot-dominated multipliers.}
\end{deluxetable}

Table~\ref{tab:pocomc_H_data} summarizes the \texttt{pocoMC} results for the hydrogen-generated synthetic data.  For the hydrogen-generated data, the helium-fit phase-channel $\chi^2$ probabilities range from $6.4 \times 10^{-2}$ to $8.5 \times 10^{-1}$, and the bolometric $\chi^2$ probabilities range from $1.5 \times 10^{-2}$ to $7.2 \times 10^{-1}$, which all remain above statistical rejection thresholds.  The correct hydrogen-fit probabilities are comparable.

Table~\ref{tab:pocomc_He_data} presents the corresponding results for the helium-generated synthetic data.  The hydrogen-fit phase-channel $\chi^2$ probabilities range from $4.2 \times 10^{-2}$ to $5.3 \times 10^{-1}$, and the bolometric $\chi^2$ probabilities range from $2.6 \times 10^{-1}$ to $9.8 \times 10^{-1}$. All fits are statistically acceptable.  The correct helium-fit probabilities are similarly acceptable for all multipliers. An incorrect atmosphere model that appears strongly disfavored at the fixed reference configuration can become statistically acceptable once the full parameter space is explored.

\begin{deluxetable}{c cccc c}
\tabletypesize{\footnotesize}
\setlength{\tabcolsep}{2pt}
\caption{Maximum-likelihood diagnostics for helium-generated synthetic data\label{tab:pocomc_He_data}}
\tablehead{
  & \multicolumn{2}{c}{He atm.\ (correct)} & \multicolumn{2}{c}{H atm.\ (incorrect)} \\[-4pt]
  \colhead{$R_{\rm synth}$}
  & \colhead{$\chi^2_{\rm pc}$ (Prob.)}
  & \colhead{$\chi^2_{\rm b}$ (Prob.)}
  & \colhead{$\chi^2_{\rm pc}$ (Prob.)}
  & \colhead{$\chi^2_{\rm b}$ (Prob.)}
  & \colhead{$\Delta\!\ln\!\mathcal{Z}$}
}
\startdata
0.061  & 2941 (.329) & 12.8 (.979) & 2942 (.325) & 12.8 (.979) & $+0.6$ \\
1.224  & 2881 (.634) & 20.7 (.710) & 2902 (.528) & 21.3 (.676) & $+12.5$ \\
2.447  & 2906 (.505) & 27.5 (.331) & 2940 (.334) & 27.7 (.323) & $+19.8$ \\
3.671  & 2986 (.152) & 27.8 (.318) & 3041 (.042) & 29.2 (.256) & $+31.7$ \\
4.894  & 2886 (.609) & 23.9 (.523) & 2929 (.388) & 24.2 (.506) & $+26.6$ \\
6.118  & 2888 (.600) & 16.8 (.888) & 2940 (.333) & 17.6 (.859) & $+30.0$ \\
\enddata
\tablecomments{$\chi^2$ values and probabilities evaluated at the maximum-likelihood \texttt{pocoMC} sample for the helium-generated data.  For $\Delta\ln\mathcal{Z} = \ln\mathcal{Z}_{\rm He} - \ln\mathcal{Z}_{\rm H}$, the positive values favor the helium model.  The evidence consistently and strongly favors the correct model ($\Delta\ln\mathcal{Z} \sim +12.5$ to $+31.7$).}
\end{deluxetable}

\subsection{Reliability of the Correct-Atmosphere Radius Posteriors}\label{subsec:calib_correct}

The acceptable $\chi^2$ values exhibited in \S\ref{subsec:sampled_fits} show that the sampler produces statistically good fits, but they do not by themselves establish that the radius posteriors are converged. Before turning to the atmospheric mismatch results in \S\ref{subsec:H_to_He} and \S\ref{subsec:He_to_H}, we confirm that the radius posteriors obtained with the correct atmosphere are converged. We verify this directly through convergence checks in which we double the effective-sample-size of the \texttt{pocoMC} runs and continue them with \texttt{emcee}. The radius posteriors are stable under both checks for all four check cases (see Appendix~\ref{app:convergence}). Since the correct-atmosphere radius posteriors are converged, any systematic displacement of the injected radius from the bulk of the posterior in the incorrect atmosphere fits is attributable to the atmospheric composition mismatch rather than to a sampling issue.

\subsection{Hydrogen Data Fit with Helium: The Dangerous Case}\label{subsec:H_to_He}
 
When hydrogen-generated synthetic data ($H,d$) are fit with a helium atmosphere model ($He,m$), the helium model's best-fit parameters found by the sampler while exploring parameter space can achieve acceptable residuals as quantified by $\chi^2$, but in the process, the radius posterior becomes significantly biased.

The bias is apparent in Figure~\ref{fig:showcase_MR} (top row), which compares the correct (blue) and incorrect (orange) atmosphere $M$--$R$ posteriors for selected multipliers.  We show $1\times$ (the baseline ratio) to illustrate the background-dominated regime where mismatch is undetectable, $60\times$ as a representative case in which the $He,m\to H,d$ bias becomes severe (incorrect model percentile $< 0.1$\%), and $100\times$ as a further spot-dominated case that confirms the persistence of the effect.  At $1\times$, the two posteriors overlap substantially, and the injected parameters lie comfortably within both.  At $60\times$ and $100\times$, the helium-fit posterior has shifted to significantly larger radii, with the injected value falling outside the $2\sigma$ contour.  The corresponding 1D radius posteriors are shown in the top row of Figure~\ref{fig:1D_radius}. The incorrect model distribution is displaced well above the injected radius.

\begin{figure*}
\centering
\includegraphics[width=\textwidth]{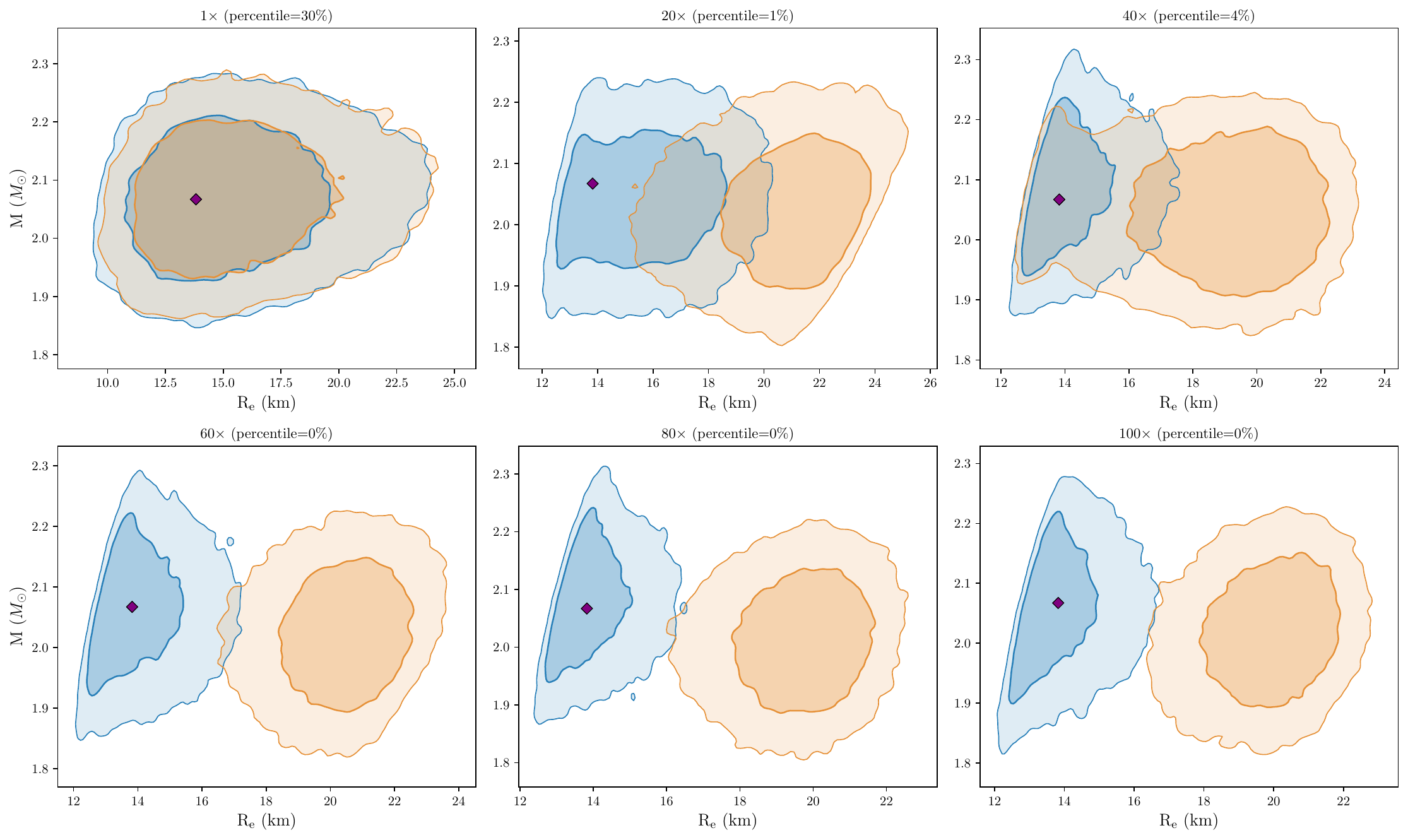}
\caption{Mass--radius posteriors for all six multipliers with hydrogen-generated synthetic data.  The blue contours show the correct H~fit, and the orange contours show the incorrect He~fit.  The purple diamond marks the injected values.  The percentile of $R_{\rm inj}$ within the correct H-fit posterior is 33rd, 19th, 45th, 51st, 51st, and 58th for $1\times$ through $100\times$, respectively, confirming that the correct model is unbiased.  The corresponding incorrect He-fit percentiles are 30th, 0.8th, 4th, 0.06th, 0.08th, and 0.07th, showing that the He-fit posterior is consistently shifted to larger radii, with $R_{\rm inj}$ falling below the $2\sigma$ threshold in most spot-dominated cases.}
\label{fig:all_MR_H}
\end{figure*}

Figure~\ref{fig:all_MR_H} shows the $M$--$R$ posteriors for all six multipliers.  The helium fit posterior is shifted to larger radii relative to the hydrogen fit posterior, and the injected radius falls deep into the low-percentile tail of the incorrect model distribution for all spot-dominated cases.

This is the more dangerous mismatch direction in the present configuration. The incorrect model yields a statistically acceptable fit while returning a significantly biased radius.
 
\subsection{Helium Data Fit with Hydrogen: Disfavored but Less Damaging}\label{subsec:He_to_H}

When helium-generated synthetic data ($He,d$) are fit with a hydrogen atmosphere model ($H,m$), the hydrogen model also achieves acceptable $\chi^2$ values after posterior sampling.  However, unlike the $He,m~\to~H,d$ case, the injected radius remains within $2\sigma$ of the incorrect model posterior for the checked multipliers (see the bottom rows of Figures~\ref{fig:showcase_MR} and \ref{fig:1D_radius}).  The incorrect model percentile values range from ${\sim}\,21\%$ at $1\times$ to ${\sim}\,97\%$ at $80\times$, but none crosses the $2\sigma$ threshold into the extreme tails.
 
Figure~\ref{fig:all_MR_He} shows the $M$--$R$ posteriors for all six multipliers in this direction.  The incorrect hydrogen-model contours (orange) are displaced to smaller radii relative to the correct helium-model contours (blue), but the displacement is modest compared to the reverse direction, and the injected parameters remain within the $2\sigma$ credible region at every multiplier.

Therefore, although the incorrect hydrogen model is statistically disfavored by the Bayesian evidence (Tables~\ref{tab:pocomc_He_data}), it does not produce the same level of radius bias observed in the $He,m~\to~H,d$ direction.
 
\subsection{Direct Comparison of the Two Mismatch Directions}\label{subsec:comparison}

\begin{figure*}
\centering
\includegraphics[width=\textwidth]{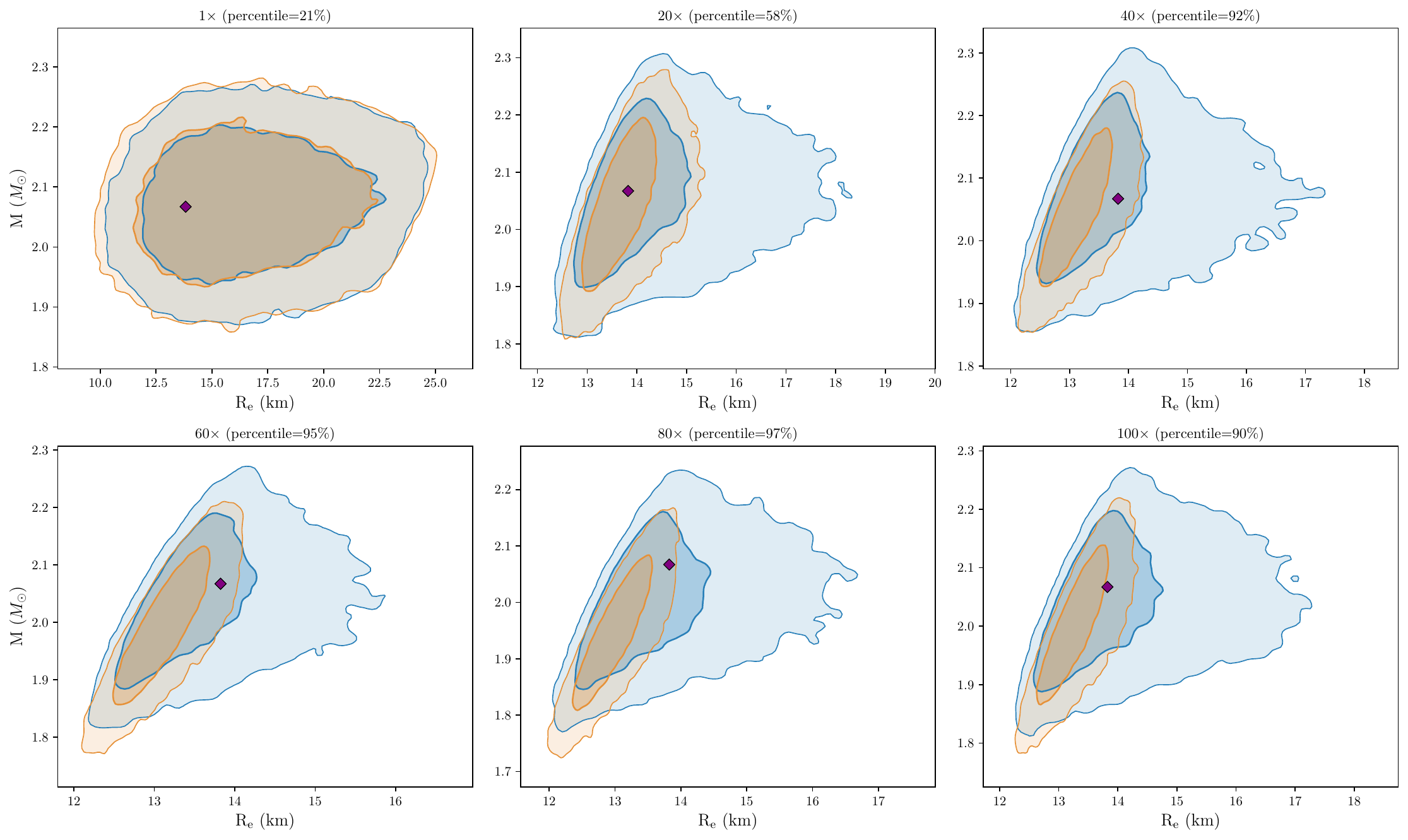}
\caption{Mass--radius posteriors for all six multipliers with helium-generated synthetic data.  The blue contours show the correct He~fit, and the orange contours show the incorrect H~fit.  The purple diamond marks the injected values.  The percentile of $R_{\rm inj}$ within the correct He-fit posterior is 18th, 43rd, 68th, 71st, 68th, and 58th for $1\times$ through $100\times$, respectively.  The corresponding incorrect H-fit percentiles are 21st, 58th, 92nd, 95th, 97th, and 90th.  The H-fit posterior is displaced to smaller radii relative to the He-fit posterior, but the injected radius remains within $2\sigma$ for all checked multipliers.}
\label{fig:all_MR_He}
\end{figure*}

The correct model percentiles suggest that the correct atmosphere recovers the injected radius without significant bias.  The asymmetry between the two incorrect model directions is clear. Fitting hydrogen-generated data with a helium model produces incorrect model percentiles at or below 6\% (and frequently below 1\%) for all spot-dominated multipliers (Figure~\ref{fig:all_MR_H}), while fitting helium-generated data with a hydrogen model produces incorrect model percentiles that remain between $\sim 21$\% and $\sim 97$\%, consistently within $2\sigma$ (Figure~\ref{fig:all_MR_He}).

The two directions share one important property. In both cases, the incorrect atmosphere model achieves statistically acceptable $\chi^2$ values after posterior sampling.  The difference lies in the consequences of that acceptable fit. The $He,m~\to~H,d$ direction produces large radius bias, while the $H,m~\to~He,d$ direction does not, in the checked J0740-like configuration.

\subsection{Bayesian Evidence as a Complement to $\chi^2$}\label{subsec:evidence_results}
 
The $\Delta\ln\mathcal{Z}$ values reported in Tables~\ref{tab:pocomc_H_data} and \ref{tab:pocomc_He_data} show that the Bayesian evidence favors the correct atmospheric model for all spot-dominated multiplers in both mismatch directions. At the realistic $1\times$ baseline, the evidence difference is small: $\Delta\ln\mathcal{Z}=-0.9$  for hydrogen-generated data fit with a helium model and $\Delta\ln\mathcal{Z}=+0.6$ for helium-generated data fit with a hydrogen model. In the former case, despite the slight preference for the incorrect atmosphere, the inferred radius is not significantly biased. Thus, in the J0740-like baseline regime, the two atmosphere models are effectively difficult to distinguish and the small evidence difference does not translate into a consequential radius bias.

For the hydrogen-generated data, $\Delta\ln\mathcal{Z}$ ranges from approximately  $+4$ to $ +23$ in favor of the hydrogen model.  For the helium-generated data, $\Delta\ln\mathcal{Z}$ ranges from approximately $+12.5$ to $ +31.7$ in favor of the helium model.

While both the correct and incorrect atmosphere models can achieve acceptable phase-channel and bolometric $\chi^2$ values, the evidence retains additional sensitivity to the atmospheric composition mismatch.  The evidence integrates over the full posterior volume, penalizing models that require fine-tuned parameter combinations to achieve their best fit.  The contrast is apparent in Tables~\ref{tab:pocomc_H_data} and \ref{tab:pocomc_He_data}. For the spot-dominated cases, rows where the correct and incorrect atmosphere $\chi^2$ probabilities are nearly identical are nevertheless accompanied by positive $\Delta\ln\mathcal{Z}$ values of approximately $+4$ or larger.

The non-monotonic behavior of $\Delta\ln\mathcal{Z}$ with increasing multiplier is a consequence of the independent Poisson realizations used to generate each synthetic dataset.  Because each multiplier corresponds to a separate realization, the specific noise pattern varies from dataset to dataset, producing fluctuations in the evidence difference that do not reflect a smooth trend.

\FloatBarrier
\section{Discussion}\label{sec:discussion}

The results presented in \S\ref{sec:results} confirm that atmospheric composition mismatch can produce biased radius estimates while remaining hidden inside a statistically acceptable fit.  The pulse-profile model has substantial freedom.  When the assumed atmosphere is incorrect, the posterior exploration can compensate for the incorrect beaming by adjusting other parameters, yielding an acceptable $\chi^2$ at the cost of a biased radius posterior.  The subsections below discuss the physical interpretation of the observed asymmetry, the insufficiency of $\chi^2$ alone, and the broader implications for NICER-like analyses.

\subsection{Why the $He,m~\to~H,d$ Direction Is More Dangerous}\label{subsec:why_H_to_He}

Our results show that fitting hydrogen-generated data with a helium model produces a larger radius bias in a J0740-like configuration. The origin of this asymmetry may lie in the high compactness of PSR~J0740$+$6620. The amount of light bending grows rapidly as a star approaches the photon sphere but changes only slowly in the Newtonian limit, meaning the bending is far more sensitive to compactness for compact stars than for extended ones. At the compactness of our reference configuration, decreasing the radius therefore increases the light bending sharply, whereas increasing the radius by the same amount reduces it only slightly. Since stronger light bending keeps the hot spot in view over a larger fraction of the rotation, the fractional modulation of the waveform varies steeply with radius below the injected value and only weakly above it. A model that needs to reduce the modulation can do so with a small decrease in radius, whereas a model that needs to increase the modulation must move to a much larger radius.

A helium beaming pattern is broader and more isotropic than the hydrogen pattern that generated the data (see \S\ref{sec:atm}), and produces too little modulation at the injected radius. To recover the modulation present in the hydrogen-generated waveform, the sampler must move toward larger radii, i.e., toward lower compactness, which is the direction in which the light bending responds weakly. A large change in $R_e$ is therefore needed, and the helium-fit posteriors are displaced to substantially larger radii, with the injected radius falling below the $2\sigma$ threshold in most spot-dominated cases.

\subsection{Why the $H,m~\to~He,d$ Direction Is Less Damaging}\label{subsec:why_He_to_H}

In the reverse direction, the fitting model starts from a beaming pattern that is more narrowly peaked than the true helium beaming, and therefore produces too much modulation at the injected radius. Correcting this requires moving toward a higher compactness. A comparatively small adjustment in $R_e$ suffices, and the incorrect hydrogen model can accommodate the mismatch without being driven into a strongly biased region of the parameter space.

The asymmetry is reflected in the widths of the radius posteriors. At the $180\times$ multiplier (Table~\ref{tab:Re_convergence}), the mismatched helium fit to hydrogen data has a $\pm 1\sigma$ radius width of $3.46$~km, roughly five times the $0.71$~km width of the mismatched hydrogen fit to helium data, and it develops a pronounced tail toward large radii. A broad, asymmetric posterior extending to large radii is expected when the waveform constrains compactness only weakly on the low-compactness side.

The geometric parameters adjust in a manner consistent with this picture. In the hydrogen-fit-to-helium runs the non-radius parameters move only modestly. The observer angle is recovered well ($\theta_{obs}=1.530$ and $1.524$ at $20\times$ and $60\times$ v.s. the injected 1.561), both spots migrate to the equator ($\theta_{c,1/2}=1.579/1.77$ at $20\times$, $1.508/1.483$ at $60\times$, near $\pi/2$ and near the line of sight), and the spot sizes shrink ($\Delta\theta_{1\&2}\approx0.071$ vs. the injected $0.087$ and $0.066$). These adjustments broaden the effective angular beaming, and the incorrect hydrogen model recovers acceptable waveforms without forcing the radius into a comparably biased region.  The radius remains within $2\sigma$ of the incorrect hydrogen-model posterior for all cases, even though the Bayesian evidence consistently favors the correct helium model.

We stress that this interpretation of the asymmetry is suggested by the results presented in \S\ref{sec:results}, not derived as a general concept.  Whether the asymmetry persists for other pulsar geometries, count levels, compactness configurations, and background conditions remains an open question.

\subsection{Why $\chi^2$ Should Be Supplemented by the Bayesian Evidence}\label{subsec:chi2_insufficient}

A central finding of this work is that a model can pass both phase-channel and bolometric $\chi^2$ checks while returning a misleading radius posterior. This is a concern because, in a real analysis, the actual atmospheric composition is unknown, meaning that given the statistically acceptable fit from $\chi^2$ alone, there is no way to tell whether the acceptable fit corresponds to the correct composition or to an incorrect one that has absorbed the mismatch through shifting other parameters. This is especially apparent in the $He,m~\to~H,d$ direction, where the $\chi^2$ values for the helium model are acceptable for every case we examined despite the radius being biased beyond $2\sigma$ in most spot-dominated cases.

This result is particularly important for analyses that aim to provide accurate equation-of-state constraints.  If $\chi^2$ acceptability were treated as sufficient validation of the atmospheric composition, one could unknowingly adopt a model that produces a significantly biased radius. \citet{2025arXiv251116759H} showed that in joint NICER/XMM-Newton analyses, even a factor-of-five underestimate of the XMM-Newton background shifts the radius posterior by only $\sim 1\sigma$, and the Bayesian evidence identified the preferred background model.  The present results show that atmospheric composition mismatch can produce a larger bias while similarly evading detection by $\chi^2$ diagnostics.

\subsection{Why the Bayesian Evidence Is Valuable}\label{subsec:evidence_value}

The Bayesian evidence retains sensitivity to atmospheric composition mismatch even when both atmosphere models produce acceptable $\chi^2$ values.  In both mismatch directions and for all spot-dominated waveforms, the evidence favors the correct atmosphere, with $\Delta\ln\mathcal{Z}$ values ranging from $\sim +4$ to $\sim +31.7$.

This sensitivity arises because the evidence integrates over the full posterior volume, rather than evaluating the fit quality only at the best-fit point.  A model that achieves an acceptable best-fit $\chi^2$ but requires a fine-tuned region of parameter space to do so will be penalized relative to a model that fits comparably well over a broader parameter range.  The evidence therefore provides information about model adequacy and should be included in the diagnostic framework for pulse-profile analyses.

\subsection{Why the Asymmetry May Not Be Universal}\label{subsec:caveat}

The asymmetry we have identified between the two mismatch directions should not be treated as a general result.  Our analysis has checked a specific configuration: a J0740$+$6620-like two-spot geometry, J0740$+$6620-like total counts ($\sim 5.5 \times 10^5$), a range of spot-to-background ratios from the realistic baseline to $100\times$ the baseline, and the particular fully ionized hydrogen and helium atmosphere models used in this work.  The degree and direction of the asymmetry may change for other pulsar geometries, different count levels, different background conditions, or if the beaming patterns of the two atmosphere models become more or less distinguishable in another regime.

The mechanism we have described above does suggest one respect in which our results may generalize. If the asymmetry arises because the light bending responds steeply to increases in compactness and weakly to decreases, then it is controlled by the compactness of the star rather than by the atmospheric properties alone. We therefore expect the two mismatch directions to become more symmetric for lower-mass, and thus less compact, neutron stars, and the asymmetry to be most pronounced for the most massive sources, of which PSR~J0740$+$6620 is a representative example.

In particular, at the actual count level and spot-to-background ratio estimated for PSR~J0740$+$6620 (the $1\times$ multiplier), fitting hydrogen-generated data with a helium model does not produce a bias in the inferred radius, and the two atmosphere models are effectively indistinguishable. This baseline result is consistent with the PSR~J0740$+$6620 analysis of \citet{2023ApJ...956..138S}, who found no significant radius change among the atmosphere models they examined. Our higher-multiplier results identify a controlled regime of larger pulsed fraction in which the atmospheric assumption becomes more consequential. The bias becomes pronounced only at higher spot-to-background ratios, where the pulsed signal contributes a larger fraction of the total counts.

\subsection{Broader Implications for NICER-like Analyses}\label{subsec:broader}

Previous work has shown that PPM radius estimates are robust against incorrect assumptions about hot spot shapes and temperature distributions \citep{2013ApJ...776...19L, 2015ApJ...808...31M, 2016EPJA...52...63M} and against substantial errors in the assumed unmodulated background \citep{2025arXiv251116759H}, provided the fit is statistically acceptable.  Unlike the systematic errors studied previously, atmospheric composition mismatch can hide inside an acceptable fit while still biasing the inferred radius.  The Bayesian evidence provides a more reliable diagnostic than $\chi^2$ for identifying this type of model inadequacy.

For pulsars with substantially higher signal-to-noise ratios than PSR~J0740$+$6620, such as PSR~J0437$-$4715, the pulsed signal constitutes a much larger number of total counts and fraction of those counts.  Our spot-to-background scaling experiment shows that as this fraction increases, the bias introduced by the incorrect atmosphere grows and the Bayesian evidence increasingly favors the correct model.  This suggests that atmospheric composition mismatch is more consequential, but also detectable, for brighter sources, and that analyses of high-signal-to-noise data should routinely compare atmosphere models using the Bayesian evidence.

In the $1\times$ spot-to-background case, our analysis shows no significant bias in either mismatch direction, because the pulsed signal is too weak relative to the background to distinguish the beaming patterns. The bias becomes apparent only at higher spot-to-background ratios.  Increasing the exposure time at the same spot-to-background ratio would reduce the statistical uncertainties without changing the fractional contribution of the pulsed signal, so the beaming mismatch would remain difficult to detect.  A significant bias at the $1\times$ spot-to-background ratio would require either substantially more data than is currently feasible or a qualitative change in the analysis, such as the inclusion of imaging data that constrain the background independently.

\FloatBarrier
\section{Conclusions}\label{sec:conclusions}

In this work we have investigated whether hydrogen and helium atmospheric composition mismatch can bias the inferred radius of a PSR~J0740$+$6620-like pulsar while remaining hidden inside a statistically acceptable fit.  We generated synthetic NICER-like pulse-profile data assuming fully ionized hydrogen and helium atmospheres and fit each synthetic dataset with both atmosphere models, varying the spot-to-background ratio while holding the total expected counts fixed at $\sim 5.5 \times 10^5$.

Our central finding is that $\chi^2$ acceptability is not a sufficient safeguard against atmospheric composition bias. In both mismatch directions, posterior sampling allows the incorrect atmosphere model to recover apparently good phase-channel and bolometric fits, yet the resulting radius posteriors can still be significantly displaced from the injected value.  This is qualitatively different from the systematics studied in prior work \citep{2013ApJ...776...19L, 2016EPJA...52...63M, 2025arXiv251116759H}, where statistically acceptable fits were reliable indicators of unbiased radii in the checked configurations.

Fitting hydrogen-generated data with a helium atmosphere model can increase the inferred radius to more than $2\sigma$ above its true value for strongly spot-dominated waveforms. Fitting helium-generated data with a hydrogen atmosphere model produces smaller shifts that remain within $2\sigma$ for all checked waveforms.  The broader beaming produced by a helium atmosphere requires more aggressive parameter compensation than the narrower beaming produced by a hydrogen atmosphere, but should not be treated as a universal result, as it may depend on the assumed geometry, count level, and atmosphere models.

At the actual count level and spot-to-background ratio of PSR~J0740$+$6620, neither mismatch direction produces a detectable bias.  However, for sources with higher pulsed fractions, such as PSR~J0437$-$4715, the present results suggest that atmospheric composition mismatch could be appreciable.

The Bayesian evidence consistently distinguishes the correct atmospheric model from the incorrect one for all spot-dominated waveforms.  This makes the Bayesian evidence an essential complement to goodness-of-fit statistics in pulse-profile analyses.

\section*{Acknowledgments} \label{sec:Ack}

This work was supported in part by NASA ADAP grant 80NSSC21K0649, and was supported by NASA through the NICER mission and the Astrophysics Explorers Program. The authors acknowledge the University of Maryland supercomputing resources (http://hpcc.umd.edu) that were made available for conducting the research reported in this paper. This work also utilized computational resources provided by the NASA Center for Climate Simulation (NCCS) Discover supercomputing facility. We are grateful to the NICER team for their ongoing efforts in instrument operation, data calibration, and analysis support. This work was performed in part at the Aspen Center for Physics, which is supported by National Science Foundation grant PHY-2210452. 
AJD was supported by NASA through the NASA Hubble Fellowship grant No. HST-HF2-51553.001, awarded by the Space Telescope Science Institute, which is operated by the Association of Universities for Research in Astronomy, Inc., for NASA, under contract NAS5-26555.

\software{pocoMC \citep{2022JOSS....7.4634K},
          emcee \citep{2013PASP..125..306F},
          Python and NumPy \citep{2007CSE.....9c..10O},
          Matplotlib \citep{2007CSE.....9...90H},
          Cython \citep{2011CSE....13b..31B},
          schwimmbad \citep{2017JOSS....2..357P},
          HEASoft \cite{2014ascl.soft08004N}}

\appendix
\twocolumngrid

\section{Sampler Configuration and Convergence of the Radius Posteriors}
\label{app:convergence}

We performed all posterior sampling with \texttt{pocoMC} (\citealt{2022MNRAS.516.1644K,2022JOSS....7.4634K}), which advances an ensemble of particles through a sequence of tempered distributions bridging the prior and the posterior, using a normalizing flow to precondition the target at each step. The two parameters that principally control sampling fidelity are the effective sample size $N_{\rm eff}$ and the number of active particles $N_{\rm act}$. $N_{\rm eff}$ sets the target effective sample size of the importance-weighted particle ensemble maintained at each iteration. A larger $N_{\rm eff}$ therefore resolves the intermediate distributions more finely and reduces the variance of both the posterior and the evidence estimate, at proportionally greater cost. $N_{\rm act}$ sets how many of those particles are advanced by MCMC at each iteration, and hence the fraction of the ensemble refreshed per step. Particles are advanced in the space of the normalizing flow using the t-preconditioned Crank--Nicolson sampler, with the flow retrained at each iteration. $N_{\rm total}$ is the number of samples in the final posterior, and $N_{\rm evidence}$ is the number of importance samples drawn from the preconditioned proposal for the final evidence estimate. We set both to $65{,}536$ for every run.
 
Our production runs used $N_{\rm eff} = 32{,}256$ and $N_{\rm act} = 4{,}032$. The convergence check described in \S\ref{app:conv_results} doubles $N_{\rm eff}$ to $64{,}512$ while holding $N_{\rm act}$ and all other settings fixed.

As an independent check on the \texttt{pocoMC} results, we continued each run with \texttt{emcee} \citep{2013PASP..125..306F}. Unlike the nested-sampling scheme, whose accuracy depends on the fidelity of a learned or assumed proposal distribution and on the adequacy of a live-point set, \texttt{emcee} is a Markov chain Monte Carlo code constructed to satisfy detailed balance with respect to the target density. Its stationary distribution is therefore the posterior by construction, and given sufficient run time it is guaranteed to produce a representative sample of the posterior. Its convergence in twelve dimensions is substantially slower than that of a preconditioned sampler, which is why we use it as a validation tool rather than as our primary sampler. However, for the same reason, a stable \texttt{emcee} continuation of a \texttt{pocoMC} run is strong evidence that the \texttt{pocoMC} posterior is not an artifact of the preconditioner or of the particle ensemble.

\subsection{Convergence results}
\label{app:conv_results}

The radius $R_e$ is the parameter of interest. We therefore verify directly that its posterior is converged. We do this in two ways. First, we repeat the \texttt{pocoMC} inference at another effective-sample-size setting, $N_{\rm eff} = 64{,}512$. Doubling the original $N_{\rm eff} = 32{,}256$ doubles the number of importance samples retained at each \texttt{pocoMC} iteration, so a sampling deficiency would appear as a disagreement between the two settings. We continue each \texttt{pocoMC} run with \texttt{emcee} and examine the ``running credible intervals" of $R_e$. At each \texttt{emcee} step we compute the median and the central $1\sigma$, $2\sigma$, and $3\sigma$ intervals of $R_e$ from all samples accumulated up to that step. If the chain has converged, these bands are flat. We perform both checks for all four atmosphere configurations (H~data/H~fit, H~data/He~fit, He~data/He~fit, He~data/H~fit) at the most demanding $180\times$ multiplier, where the pulsed signal is strongest.

\begin{figure*}[!htbp]
\centering

\includegraphics[width=\textwidth]{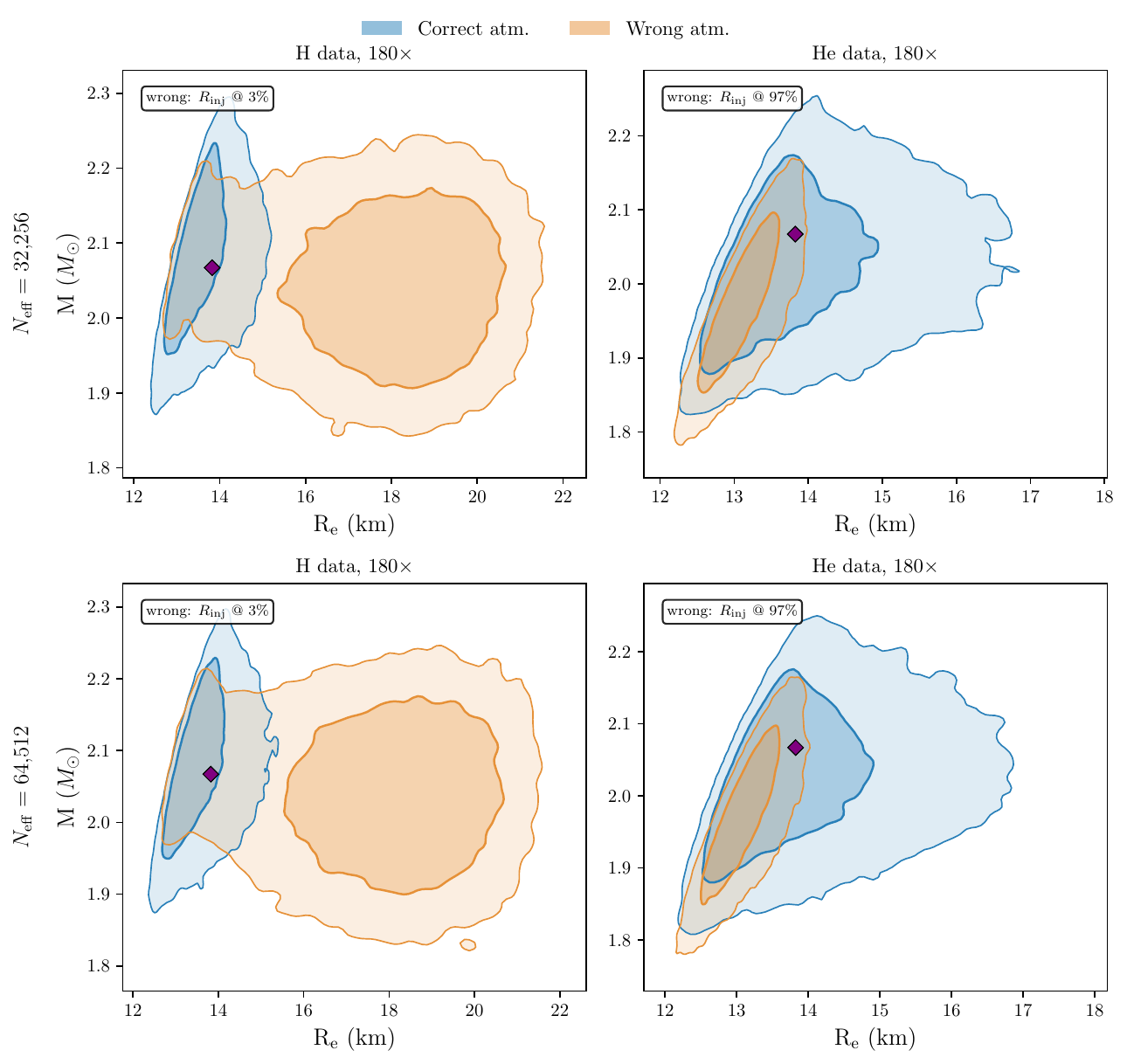}
\caption{Two-dimensional $R_e$--$M$ posteriors for $N_{\rm eff} = 32{,}256$ (top row) and $N_{\rm eff} = 64{,}512$ (bottom row) at the $180\times$ multiplier. \textit{Left:} hydrogen-generated data, with the correct H fit and the mismatched He fit overlaid. \textit{Right:} helium-generated data, with the correct He fit and the mismatched H fit overlaid. In each panel, the correct-atmosphere fit is shown in blue and the mismatched fit in orange; inner and outer contours are the $68\%$ and $95\%$ credible regions, and the purple diamond marks the injected value $(R_{\rm inj}, M_{\rm inj}) = (13.82~{\rm km}, 2.07~M_\odot)$. The radius bias of the mismatched fits, the large-radius tail in the H~data/He~fit case, and the compressed radius posterior in the He~data/H~fit case are reproduced identically at both $N_{\rm eff}$ settings.}
\label{fig:Re_M_convergence}
\end{figure*}

\begin{figure*}[!htbp]
\centering

\includegraphics[width=\textwidth]{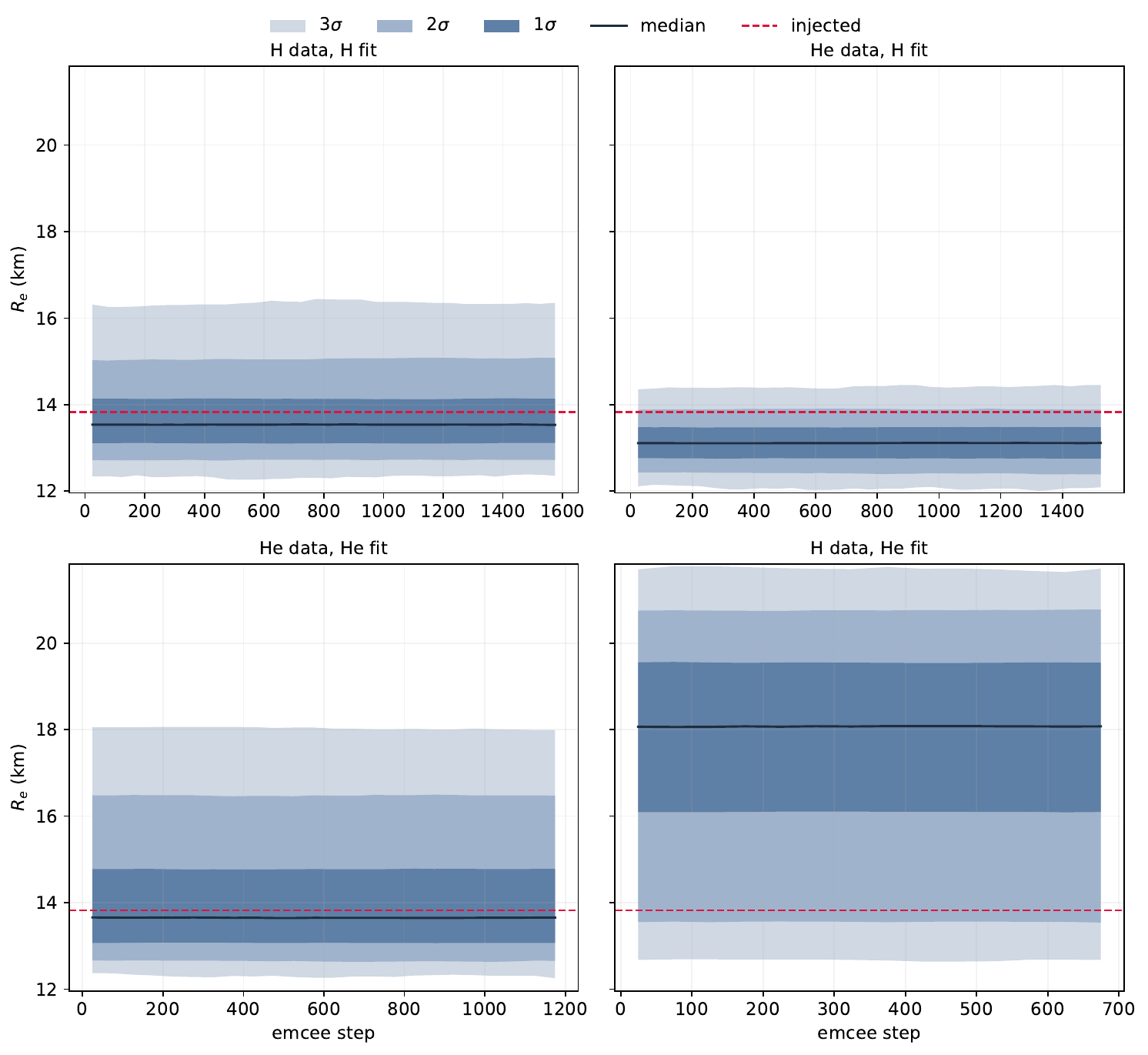}
\caption{Running credible intervals of the radius $R_e$ from \texttt{pocoMC} $N_{\rm eff} = 32{,}256$ setting, using \texttt{emcee}, for the $180\times$ multiplier for the four atmosphere configurations. The shaded bands show the running $1\sigma$, $2\sigma$, and $3\sigma$ credible intervals, the solid line the running median, and the dashed line the injected radius. The radius posterior is stable over the full chain in every configuration, confirming convergence of the \texttt{pocoMC} results.}
\label{fig:Re_trace32}
\end{figure*}

\begin{figure*}[!htbp]
\centering

\includegraphics[width=\textwidth]{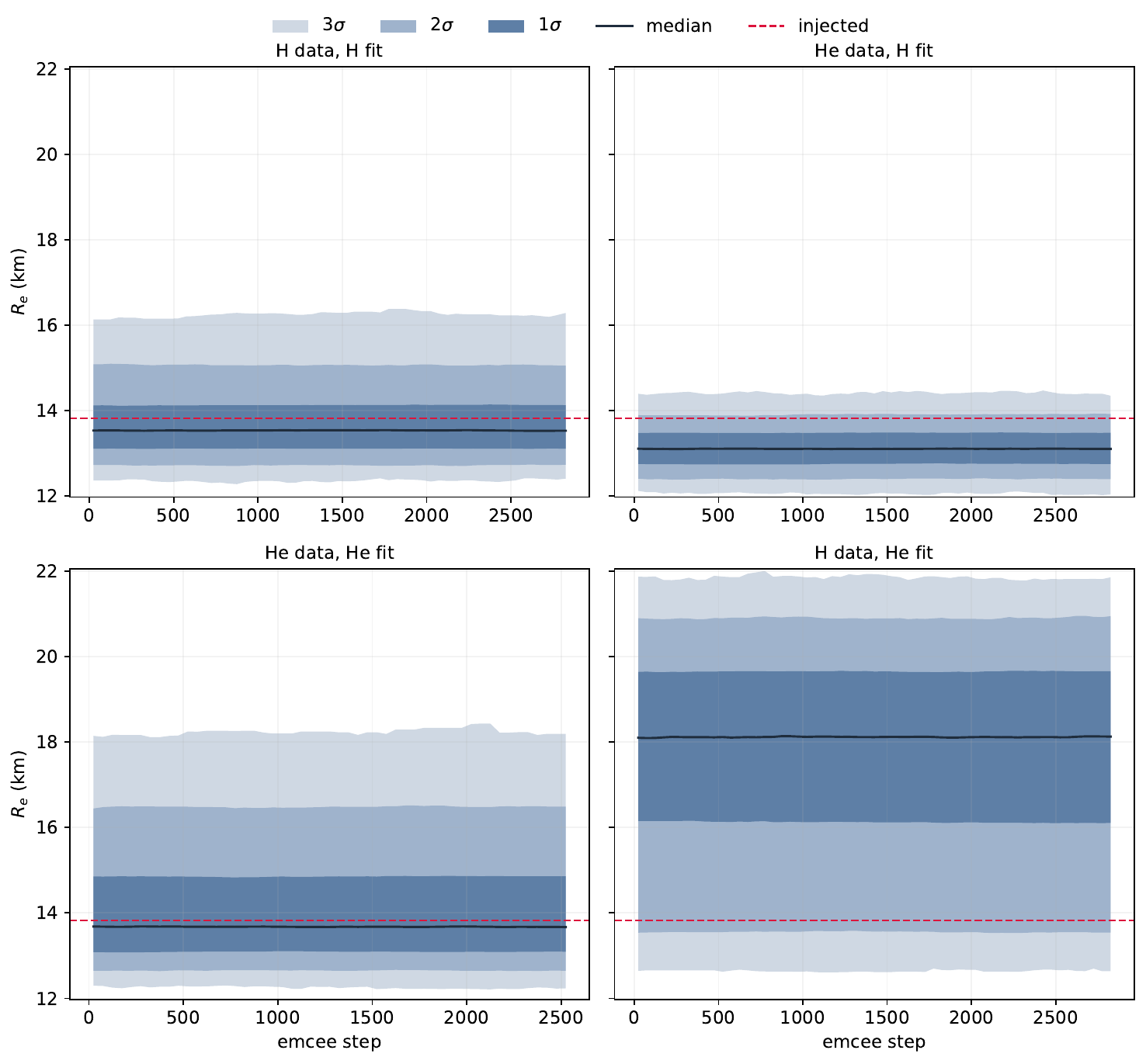}
\caption{Running credible intervals of the radius $R_e$ from \texttt{pocoMC} $N_{\rm eff} = 64{,}512$ setting, using \texttt{emcee}, for the $180\times$ multiplier for the four atmosphere configurations. The shaded bands show the running $1\sigma$, $2\sigma$, and $3\sigma$ credible intervals, the solid line the running median, and the dashed line the injected radius. The radius posterior is stable over the full chain in every configuration, confirming convergence of the \texttt{pocoMC} results.}
\label{fig:Re_trace64}
\end{figure*}

Table~\ref{tab:Re_convergence} compares the radius posterior at the two $N_{\rm eff}$ settings. For all four configurations, the radius median shifts by at most $0.064$~km when $N_{\rm eff}$ is doubled, and the $\pm1\sigma$ and $\pm2\sigma$ interval widths change by no more than a few percent, with no systematic direction. The quantile of the injected radius within the posterior is unchanged between the two settings in all four cases. Figure~\ref{fig:Re_M_convergence} shows the joint posteriors at $N_{\rm eff} = 32{,}256$ and the $N_{\rm eff} = 64{,}512$ with contours that overlie the correct and incorrect models. The strong radius bias in the mismatched fits, the asymmetric tail toward large radii in the H~data/He~fit case, and the compressed radius posterior in the He~data/H~fit case are all reproduced at both settings. The \texttt{emcee} running credible intervals, shown in Figures~\ref{fig:Re_trace32} and \ref{fig:Re_trace64}, remain stable over the full chain for every configuration. We conclude that the radius posteriors are converged at our adopted sampler settings.

\begin{deluxetable}{cccccc}
\caption{Convergence of the $R_e$ posterior in km after doubling $N_{\rm eff}$.}
\setlength{\tabcolsep}{2pt}
\tablehead{
\colhead{Configuration} & \colhead{$N_{\rm eff}$} & \colhead{Median} & \colhead{$\pm1\sigma$ width} & \colhead{$\pm2\sigma$ width} & \colhead{$\Delta$ median}
}
\startdata
\multirow{2}{*}{H data, H fit}     & $32{,}256$ & $13.532$ & $1.010$ & $2.293$ & \multirow{2}{*}{$+0.000$} \\
                                             & $64{,}512$ & $13.532$ & $1.036$ & $2.352$ & \\
\hline
\multirow{2}{*}{H data, He fit}   & $32{,}256$ & $18.080$ & $3.461$ & $7.225$  & \multirow{2}{*}{$+0.064$} \\
                                             & $64{,}512$ & $18.144$ & $3.464$ & $7.287$  & \\
\hline
\multirow{2}{*}{He data, He fit}   & $32{,}256$ & $13.666$ & $1.733$ & $3.737$ & \multirow{2}{*}{$+0.006$} \\
                                             & $64{,}512$ & $13.672$ & $1.752$ & $3.841$ & \\
\hline
\multirow{2}{*}{He data, H fit}   & $32{,}256$ & $13.112$ & $0.712$ & $1.454$ & \multirow{2}{*}{$-0.003$} \\
                                             & $64{,}512$ & $13.109$ & $0.721$ & $1.480$ & \\
\enddata
\tablecomments{The radius median and the $\pm1\sigma$ and $\pm2\sigma$ interval widths at each $N_{\rm eff}$ setting are shown in each column. Medians and widths are in km. $\Delta$ median is $\mathrm{med}_{64}-\mathrm{med}_{32}$. In every configuration, the median shift is far smaller than the $\pm1\sigma$ width, indicating that the radius posterior has converged.}
\label{tab:Re_convergence}
\end{deluxetable}

\clearpage

\bibliography{sample631}{}
\bibliographystyle{aasjournal}

\end{document}